\documentclass[a4paper,11pt]{article}
\pdfoutput=1 % if your are submitting a pdflatex (i.e. if you have images in pdf, png or jpg format)
\usepackage{jheppub}	% for details on the use of the package, please see the JHEP-author-manual
\usepackage[T1]{fontenc}
\usepackage[english]{babel}
\usepackage{makeidx}
\usepackage{amsfonts}
\usepackage{mathrsfs}
\usepackage{empheq}
\usepackage{braket}
\usepackage{tensor}
\usepackage{slashed}	
\usepackage{tikz}
\usepackage[autostyle]{csquotes}

\newcommand{\lf}{\left}				\newcommand{\rg}{\right}

\newcommand{\bea}{\begin{eqnarray}}
\newcommand{\eea}{\end{eqnarray}}
\def\nn{\nonumber}

\def\a{\alpha}		\def\b{\beta}		\def\g{\gamma}		\def\d{\delta}
\def\e{\varepsilon}			\def\h{\eta}			
				\def\l{\lambda}		\def\m{\mu}
\def\n{\nu}						\def\p{\pi}			\def\r{\rho}	
\def\s{\sigma}		\def\t{\tau}			\def\f{\phi}			\def\y{\psi}
\def\c{\chi}					\def\vf{\varphi}		

\def\G{\Gamma}						\def\L{\Lambda}

\newcommand{\calA}{{\mathcal A}}

	\newcommand{\calN}{{\mathcal N}}

\newcommand{\<}{\langle}			\renewcommand{\>}{\rangle}

	\newcommand{\setZ}{\mathbb{Z}}	
\newcommand{\setC}{\mathbb{C}}

\newcommand{\der}{{\partial}}										% Substitute \partial
\DeclareMathOperator{\re}{Re}
\DeclareMathOperator{\im}{Im}

\newcommand{\aD}{{\dot{\alpha}}}

\newcommand{\yB}{{\bar{\psi}}}
\newcommand{\cB}{{\bar{\chi}}}

\newcommand{\kslash}{{\slashed{k}}}

\newcommand{\ap}{{\alpha^\prime}}
\newcommand{\gYM}[1]{g_{\scriptscriptstyle{\text{YM}},\scriptstyle{#1}}}

\newcommand{\ie}{{\it i.e.\ }}

\newcommand{\Fab}[4]{\tensor[_2]{F}{_1}(#1,#2;#3;#4)}

\title{Yukawa's of light stringy states}

\author[a]{Pascal Anastasopoulos,}
\author[b]{Massimo Bianchi,}
\author[b]{Dario Consoli}
\affiliation[a]{Technische Univ. Wien Inst. f\"ur Theoretische Physik, A-1040 Vienna, Austria}
\affiliation[b]{Dipartimento  di  Fisica,  Universit\`a  di  Roma  ``Tor  Vergata''  \& \\ I.N.F.N.  Sezione  di  Roma  ``Tor  Vergata'', Via della Ricerca Scientifica, 00133 Roma, Italy}

\emailAdd{pascal@hep.itp.tuwien.ac.at}
\emailAdd{massimo.bianchi@roma2.infn.it}
\emailAdd{dario.consoli@roma2.infn.it}

\date{}
\abstract{Light massive string states can appear at D-brane intersections with small angles.  We compute tri-linear Yukawa couplings of such open-string states to massless ones and to one another. Due to ambiguities in the normalisation of the vertex operators, that involve twist fields, we proceed via factorization of appropriate scattering amplitudes. Some peculiar features are observed that may lead to interesting signatures at colliders in the future.}
\preprint{TUW-16-20 \\ \vspace{-8mm} \begin{flushright} ROM2F-2016-08 \end{flushright}}
 \clearpage

\begin{document}

\maketitle
\flushbottom

\section{Introduction}\label{sect:Introduction}

Orientifold compactifications is a very succesful framework for semi-realistic model building\footnote{For original work on orientifolds, see \cite{Sagnotti:1988uw, Bianchi:1988fr, Bianchi:1988ux, Pradisi:1988xd, Bianchi:1989du, Dai:1989ua, Bianchi:1990yu, Bianchi:1990tb, Bianchi:1991rd, Bianchi:1991eu,Angelantonj:1996uy,Polchinski:1995mt, Berkooz:1996km} and for recent reviews on D-brane model building, see \cite{Blumenhagen:2005mu, Blumenhagen:2006ci, Marchesano:2007de, Cvetic:2011vz, Ibanez:2012zz} and references therein.}. 
In this class of models the gauge symmetry does live on the world-volume of lower-dimensional hyperplanes, called D-branes, whereas the chiral matter is localised at the intersection of different D-brane stacks. 
Thus, all the Standard Model fields can be descrived by  open strings with either both of their ends on the same stack (gluons, SU(2) gauge fields and the Hypercharge) or on different stacks of D-branes (quarks, leptons and the Higgs)\footnote{For original work on local D-brane configurations, see \cite{Antoniadis:2000ena,Aldazabal:2000sa}. For a systematic analysis of local D-brane configurations, see \cite{Gmeiner:2005vz,Anastasopoulos:2006da,Cvetic:2009yh,Cvetic:2010mm}.}.

One of the most exciting property of this class of models is that they allow for a very low string tension scale, even of order of a few TeV's \cite{ArkaniHamed:1998rs, Antoniadis:1997zg, Antoniadis:1998ig}, therefore, stringy effects become viable candidates for physics beyond the Standard Model (see or example anomalous $Z'$ \cite{Kiritsis:2002aj, Antoniadis:2002cs, Ghilencea:2002da, Anastasopoulos:2003aj, Anastasopoulos:2004ga, Burikham:2004su, Coriano':2005js, Anastasopoulos:2006cz, Anastasopoulos:2008jt, Armillis:2008vp, Fucito:2008ai,  Anchordoqui:2011ag, Anchordoqui:2011eg, Anchordoqui:2012wt}, 
Kaluza Klein states \cite{Dudas:1999gz, Accomando:1999sj, Cullen:2000ef, Burgess:2004yq, Chialva:2005gt, Cicoli:2011yy, Chialva:2012rq},
or purely stringy signatures \cite{Bianchi:2006nf, Anchordoqui:2007da, Anchordoqui:2008ac, Lust:2008qc, Anchordoqui:2008di, Anchordoqui:2009ja, Lust:2009pz, Anchordoqui:2009mm,  Anchordoqui:2010zs, Feng:2010yx, Dong:2010jt, Carmi:2011dt, Hashi:2012ka, Anchordoqui:2014wha}\footnote{For recent reviews, see \cite{Lust:2013koa, Berenstein:2014wva}.}. In this paper we will take a different direction.

A string living between two different stacks of branes can only vibrate with frequences which are proportinal to the angle between the branes $\theta$. That generates a whole tower of massive copies, with masses proportinal to $\sqrt{\theta} M_s$, of the lowest/massless mode living at this intersection. Notice that the first excited mode will be ligher than the standard Regge excitations by a factor $\theta/\pi$.
As a concequence, massive copies for all matter fields of the Standard Model is a direct prediction of all D-brane realizations of the Standard Model. If the string scale $M_s$ is at a few TeV range, such copies can be very light (aka $light$ $stringy$ $states$) and therefore they might be visible at LHC \cite{Anastasopoulos:2011hj, Anastasopoulos:2011gn, Anastasopoulos:2014lpa, Anastasopoulos:2015dqa}. 

With this in mind, two of the present authors (A.~P. and M.~B.) have proposed to interpret the 750 GeV di-photon excess observed at LHC in terms of a massive replica of the Higgs boson and predicted the existence of a second replica at around 1053 GeV \cite{Anastasopoulos:2016cmg}. Notwithstanding the present trend to ascribe the 750 GeV excess to a fluctuation of the background rather than to the true signal of a new particle, it is interesting to analyse the couplings of light massive string states with one another and with the massless ones that should account for the standard model particles in models with open and unoriented strings and be prepared for new data coming from the LHC.

Our goal is the computation of Yukawa couplings of massive light string states to massless ones and to one another. Due to ambiguities in the normalisation of the vertex operators, we prefer to proceed via factorizations of appropriate scattering amplitudes. In particular, the knowledge of the gauge couplings and the normalizations of vertex operators and twist-field correlators is required. The gauge couplings plays a key role to fix the normalization of vertex operators and twist-field correlators that appear in amplitudes with gauge bosons in the external states or propagating in intermediate channels. Once these normalizations are fixed, we extract some of the desired Yukawas via factorization. Other Yukawa's must be extracted from amplitudes without gauge bosons, in this case the previously computed/known Yukawa's  replace the gauge couplings in order to fix normalizations.

The main difficulties that one has to face are the proper identification of BRST invariant vertex operators for massive replicas, which has largely been solved in \cite{Anastasopoulos:2016cmg} and the subsequent computation of amplitudes that involved excited twist fields, which has been addressed preliminarily in \cite{Anastasopoulos:2016cmg} based on previous work on the subject \cite{Anastasopoulos:2013sta}. 

Aim of the present paper is to fill in some of the details and compute gauge and Yukawa couplings of light massive open strings living at D-brane intersections. Although we keep open the possibility of small angles, the results will be valid in more general settings. We will also illustrate compatibility of our results with SUSY Ward identities.

The plan of the paper is as follows.
In Section \ref{sect:setupandVOs}, we will describe the set-up of intersecting D6-branes on tori and briefly recall the expressions for both massless and massive BRST invariant vertex operators (VO's). 
In Section  \ref{sect:scattering_amplitudes} we will start computing scattering amplitudes on the disk (tree level). Normalization problems will be solved by first considering amplitudes that expose vector boson exchange on one or both channels ($s$ and $t$). 
Our results will be summarised in Section \ref{sect:FinResults}.
Various appendices discuss the state-operator correspondence, the procedure to derive open-string twist field correlators and scattering amplitudes from closed-string ones, limits and identities for hypergeometric functions used to derive Yukawa couplings.

\section{Setup and vertex operators}\label{sect:setupandVOs}

Intersecting and/or magnetised D-brane configurations represent a very simple yet effective setting to embed the Standard Model or a (supersymmetric) generalisation. For simplicity we will consider intersecting D6-branes wrapping factorizable 3-cycles on a six-torus $T^6=T^2\times T^2\times T^2$. In order to compute gauge and Yukawa couplings of light massive string states, we consider three stacks labelled by $a$, $b$ and $c$.  In each torus $T^2_I$ ($I=1,2,3$) two generic stacks, let us say $a$ and $b$,  intersecting at an angle $\pi a_{ab}^I$. The intersection is supersymmetric if the condition
\begin{equation}
\pm a_{ab}^1\pm a_{ab}^2 \pm a_{ab}^3 = 0 \mod 2
\end{equation}
is satisfied for some choice of signs\footnote{Semi-realistic MSSM constructions on factorizable orbifolds can be found in \cite{Blumenhagen:2000wh,Angelantonj:2000hi,Aldazabal:2000cn, Aldazabal:2000dg, Forste:2000hx, Ibanez:2001nd, Cvetic:2001tj, Cvetic:2001nr, 
 Honecker:2003vq, Cvetic:2004nk, Honecker:2004np, Honecker:2004kb, Blumenhagen:2004xx, Blumenhagen:2005tn, Gmeiner:2005vz, Bailin:2006zf, 
 Cvetic:2006by, Chen:2007px, Bailin:2007va,Gmeiner:2007zz, Bailin:2008xx, Gmeiner:2008xq, Honecker:2012qr, Honecker:2013kda}.}.

In order to have some non-vanishing Yukawa couplings, since the three stacks of branes form a triangle in each subtorus, we take
\begin{align}
\label{eq:non_vanish_Yukawas_consitions}
a_{ab}^1+a_{bc}^1+a_{ca}^1&=0  \\
a_{ab}^2+a_{bc}^2+a_{ca}^2&=0  \\
a_{ab}^3+a_{bc}^3+a_{ca}^3&=-2 
\end{align}
Moreover in order to preserve the same SUSY at the various internal dimensions one has
\begin{align}
\label{eq:susy_conditions}
a_{ab}^1+a_{ab}^2+a_{ab}^3 &=0\quad \qquad 0<a_{ab}^1<1 \,\qquad 0<a_{bc}^1<1 \quad -1<a_{ca}^1<0 \\
a_{bc}^1+a_{bc}^2+a_{bc}^3 &=0 \quad \qquad 0<a_{ab}^2<1 \,\qquad 0<a_{bc}^2<1 \quad -1<a_{ca}^2<0\\
a_{ca}^1+a_{ca}^2+a_{ca}^3 &=-2 \quad -1<a_{ab}^3<0 \quad -1<a_{bc}^3<0 \quad -2<a_{ca}^3<0
\end{align}

\begin{figure} [t]
\centering
\begin{tikzpicture}
%\draw [step=0.25] (0,0) grid (8,4);
\draw [dashed,gray] (0,-0.2) -- (0,4.2) (4,-0.2) -- (4,4.2) (8,-0.2) -- (8,4.2);
\draw [red] (-0.2,0) -- (8.2,0) (-0.2,2) -- (8.2,2) (-0.2,4) -- (8.2,4);
\draw [blue] (45:{-0.2*1.4}) -- (4,4)--++(45:{0.2*1.4}) (2,0)++(45:{-0.2*1.4}) -- (6,4)--++(45:{0.2*1.4}) (4,0)++(45:{-0.2*1.4}) -- (8,4)--++(45:{0.2*1.4})
(0,2)++(45:{-0.2*1.4}) -- (2,4)--++(45:{0.2*1.4}) (6,0)++(45:{-0.2*1.4}) -- (8,2)--++(45:{0.2*1.4})
(0,4)++(45:{-0.2*1.4})--++(45:{0.4*1.4}) (8,0)++(45:{-0.2*1.4})--++(45:{0.4*1.4});
\draw [green] (1,-0.2) -- (1,4.2) (5,-0.2) -- (5,4.2);
\fill (0,0) circle (2 pt) (4,0) circle (2 pt) (8,0) circle (2 pt)
(0,2) circle (2 pt) (4,2) circle (2 pt) (8,2) circle (2 pt)
(0,4) circle (2 pt) (4,4) circle (2 pt) (8,4) circle (2 pt);
\fill (-2 pt+2 cm,-2 pt) rectangle (2 pt+2 cm,2 pt) (-2 pt+2 cm,-2 pt+2cm) rectangle (2 pt+2 cm,2 pt+2cm)
(-2 pt+2 cm,-2 pt+4cm) rectangle (2 pt+2 cm,2 pt+4cm)
(-2 pt+6 cm,-2 pt) rectangle (2 pt+6 cm,2 pt) (-2 pt+6 cm,-2 pt+2cm) rectangle (2 pt+6 cm,2 pt+2cm)
(-2 pt+6 cm,-2 pt+4cm) rectangle (2 pt+6cm,2 pt+4cm);
\draw (-2pt+1cm,-2pt) rectangle (2pt+1cm,2pt) (-2pt+5cm,-2pt) rectangle (2pt+5cm,2pt)
(-2pt+1cm,2cm-2pt) rectangle (2pt+1cm,2cm+2pt) (-2pt+5cm,2cm-2pt) rectangle (2pt+5cm,2cm+2pt)
(-2pt+1cm,4cm-2pt) rectangle (2pt+1cm,4cm+2pt) (-2pt+5cm,4cm-2pt) rectangle (2pt+5cm,4cm+2pt);
\draw (1,1) circle (2pt) (1,3) circle (2pt) (5,1) circle (2pt) (5,3) circle (2pt);
\end{tikzpicture}
\caption{A simple configuration of three stacks of D6-branes in a torus $T^2_I$. The angles in the figure are large for illustrative purposes. The fundamental cell of the torus is delimited by the intersections represented with filled circles.}
\label{fig:3_intersecting_branes}
\end{figure}
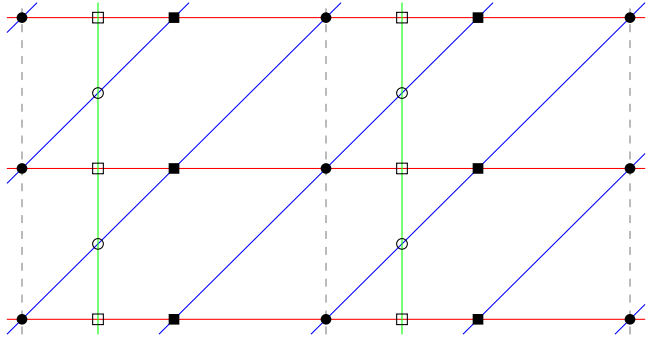

Strings ending on two different stacks of branes give rise to a massless (chiral) spectrum with multiplicity given by the number of intersections and a massive spectrum. In \cite{Anastasopoulos:2016cmg} a detailed analysis of the spectrum has been performed, relying on the helicity super trace. The explicit expression of the BRST invariant vertex operators for the first few massive states was found and, in the supersymmetric case, the structure of the super-multiplets was identified, including the relevant SUSY transformations.

We will not explicitly display here the expressions of the vertex operators for (massive) spinor and scalar states whose interactions we would like to study later on. For the sake of self-containedness, we only present some of them here 
%{\bf Notice that the VO's are angle dependant and we present them }
\bea
&& V_{\f_0=\f_0^{ab}}^{(-1)} =C_{\f_0} e^{-\f_{10}} \f_0 e^{-\vf} \s_{a^1_{a,b}} \s_{a^2_{a,b}} \s_{1+a^3_{a,b}}
e^{i[a^1_{a,b}\vf_1+a^2_{a,b}\vf_2+(a^3_{a,b}+1)\vf_3]} e^{i k X} \qquad ~~~ \\
&& V_{\y_0=\c_0^{bc}}^{(-\frac{1}{2})} =C_{\y_0} e^{-\f_{10}} \y_0^\a S_\a e^{-\frac{\vf}{2}} \s_{a^1_{b,c}} \s_{a^2_{b,c}} \s_{1+a^3_{b,c}} e^{i[(a^1_{b,c}-\frac{1}{2})\vf_1+(a^2_{b,c}-\frac{1}{2})\vf_2+(a^3_{b,c}+\frac{1}{2})\vf_3]} e^{i k X}\qquad ~~~ 
\\
&& V_{\c_0=\c_0^{ca}}^{(-\frac{1}{2})}=C_{\c_0} e^{-\f_{10}} \c_0^\a S_\a e^{-\frac{\vf}{2}} \s_{1+a^1_{c,a}} \s_{1+a^2_{c,a}} \s_{1+a^3_{c,a}} e^{i[(a^1_{c,a}+\frac{1}{2})\vf_1+(a^2_{c,a}+\frac{1}{2})\vf_2+(a^3_{c,a}+\frac{1}{2})\vf_3]} e^{i k X}\qquad ~~~ \\
%&& V_{\y_1=\c_1^{bc}}^{(-\frac{1}{2})}=C_{\y_1} e^{-\f_{10}} \y_1^\a S_\a e^{-\frac{\vf}{2}} \t_{a^1_{b,c}} \s_{a^2_{b,c}} \s_{1+a^3_{b,c}} e^{i[(a^1_{b,c}-\frac{1}{2})\vf_1+(a^2_{b,c}-\frac{1}{2})\vf_2+(a^3_{b,c}+\frac{1}{2})\vf_3]} e^{i k X}\qquad ~~~ \nn\\
&& V_{\y_1=\c_1^{bc}}^{(-\frac{1}{2})}=C_{\y_1} e^{-\f_{10}} \y_1^\a S_\a e^{-\frac{\vf}{2}} \t_{a^1_{b,c}} \s_{a^2_{b,c}} \s_{1+a^3_{b,c}} e^{i[(a^1_{b,c}-\frac{1}{2})\vf_1+(a^2_{b,c}-\frac{1}{2})\vf_2+(a^3_{b,c}+\frac{1}{2})\vf_3]} e^{i k X}\qquad ~~~ \nn\\
&&~~~~~~~~~~~~+C_{\tilde\y_1} e^{-\f_{10}} \tilde\y^\dagger_{1\dot\a} C^{\dot\a} e^{-\frac{\vf}{2}} \s_{a^1_{b,c}} \s_{a^2_{b,c}} \s_{1+a^3_{b,c}} e^{i[(a^1_{b,c}+\frac{1}{2})\vf_1+(a^2_{b,c}-\frac{1}{2})\vf_2+(a^3_{b,c}+\frac{1}{2})\vf_3]} e^{i k X}\qquad ~~~ \\
&& V_{\c_1=\c_1^{ca}}^{(-\frac{1}{2})}=C_{\c_1} e^{-\f_{10}} \c_1^\a S_\a e^{-\frac{\vf}{2}} \s_{1+a^1_{c,a}} \s_{1+a^2_{c,a}} \t_{1+a^3_{c,a}} e^{i[(a^1_{c,a}+\frac{1}{2})\vf_1+(a^2_{c,a}+\frac{1}{2})\vf_2+(a^3_{c,a}+\frac{1}{2})\vf_3]} e^{i k X}\qquad ~~~ \nn\\
&&~~~~~~~~~~~~+C_{\tilde\c_1} e^{-\f_{10}} \tilde\c^\dagger_{1\dot\a} C^{\dot\a} e^{-\frac{\vf}{2}} \s_{1+a^1_{c,a}} \s_{1+a^2_{c,a}} \s_{1+a^3_{c,a}} e^{i[(a^1_{c,a}+\frac{1}{2})\vf_1+(a^2_{c,a}+\frac{1}{2})\vf_2+(a^3_{c,a}-\frac{1}{2})\vf_3]} e^{i k X}\qquad ~~~ 
\eea
The masses of the corresponding fields are:
\bea
m^2_{\f_0}= 0 ~~&,&~~~~~ m^2_{\y_0}= 0     ~~~~~~~~,~~~~~ m^2_{\c_0}= 0 ~\\
                       ~~&,&~~~~~ m^2_{\y_1}= a^1_{bc} /\a'~~,~~~~~ m^2_{\c_1}= (1-|a^3_{ca}|) /\a' ~~.
\eea
The rest can be found in Appendix \ref{States and vertex operators} or in \cite{Anastasopoulos:2016cmg}\footnote{For a detailed discussion on vertex operators of massless states for arbitrary intersection angles, see \cite{Cvetic:2006iz,Bertolini:2005qh}, for a generalization to massive states, see \cite{Anastasopoulos:2011hj}, and for a discussion on instantonic states
at the intersection of D-instanton and D-brane at arbitrary angles, see \cite{Cvetic:2009mt}.}. Notice that vertex operators involve twist fields and their excitations, that generalise standard twist-fields present in orbifold models \cite{Dixon:1985jw, Dixon:1986jc, Dixon:1986qv}. While the latter depend on rational twists ($\theta =k/n$), open string twist-fields depend on continuous variables, such as the angles of the intersections or, in T-dual contexts, the magnetic fluxes.

We would like to briefly review our strategy to derive open-string twist field correlators from closed-string ones. The procedure we adopt is essentially based on the systematic construction of open string theories starting from closed-string theories that admit an anti-holomorphic involution exchanging Left- and Right- movers \cite{Anastasopoulos:2011gn}. 

In general $n$-pt closed-string correlators on the sphere depend on $n-3$ conformal invariant complex cross-ratios. Similarly $n$-pt open-string correlators on the disk depend on $n-3$ conformal invariant real cross-ratios\footnote{We are implicitly identifying the unit disk with the upper half plane whose boundary, where open strings are inserted, corresponds to $\im(z) =0$.}.

The simplest non-trivial case is $n=3$. In both cases the correlator is a constant (OPE coefficient) independent of the positions of the insertion points that can be taken to be $z_1=0,z_2=1, z_3=\infty$. The overall normalisation is notoriously difficult to determine and it turns out to be simpler to derive tri-linear couplings factorizing 4-pt amplitudes that can be normalised unambiguously at least in one channel, where gauge fields or gravitons are exchanged. 

Let us then focus on 4-pt correlation functions. In the closed-string case one has
\begin{align}
{\cal G}_4(z_i, \bar{z}_i) &= \langle \Phi_{h_1, \bar{h}_1}(z_1, \bar{z}_1) \Phi_{h_2, \bar{h}_2}(z_2, \bar{z}_2) \Phi_{h_3, \bar{h}_3}(z_3, \bar{z}_3) \Phi_{h_4, \bar{h}_4}(z_4, \bar{z}_4)\rangle \nn\\
&= \prod_{i<j} |z_{ij}|^{- h_i-\bar{h}_i- h_j-\bar{h}_j + {\Delta + \frac{\bar\Delta}{3}}} \sum_{h,\bar{h}} {\cal F}_h(z) \bar{\cal F}_{\bar{h}}(\bar{z})
\end{align}
where $z_{ij} =z_i - z_j$, $z=z_{12} z_{34}/z_{14}z_{23}$, $\Delta = h_1+h_2+h_3+h_4$ and ${\cal F}_h(z)$ are the conformal blocks, with $h$ running over the primary fields that appear in the OPE of both
$\Phi_{h_1, \bar{h}_1}$ with $\Phi_{h_2, \bar{h}_2}$ and of $ \Phi_{h_3, \bar{h}_3}$ with $ \Phi_{h_4, \bar{h}_4}$. Up to an overall factor $z^{h-h_1-h_2}$ the conformal blocks admit a power series expansion in the cross-ratio $z$, pretty much as the characters $\chi_h(q) = {\rm Tr}_{{\cal H}_h} q^{L_0 - {\frac{c}{24}}}$ that appear at one-loop (torus) admit a power series expansion in $q=e^{2\p i \t}$.

The relation between 4-pt amplitudes at tree-level and one-loop vacuum amplitudes is more than an analogy. Indeed most of the computations of correlators with non-trivial monodromies on the sphere can be holomorphiclly mapped into one loop amplitudes on a covering torus, possibly a `fake' one \cite{Dixon:1986qv}, with $\tau=\tau(z)$ (holomorphic map!)
\begin{equation}
{\cal G}^\textup{tree}_4(z, \bar{z}) = {\cal Z}^\textup{1-loop}_0(\tau, \bar{\tau})
\end{equation}
Sometimes the latter can be conveniently factorized into a `quantum' part ${\cal Z}_\textup{qu}(\tau, \bar{\tau})$ and a `classical' part  ${\cal Z}_\textup{cl}(\tau, \bar{\tau}) = \sum_I e^{-S_\textup{cl}(\tau, \bar{\tau}; I)}$ where $I$ labels the `instantons' to be summed over.

Exploiting the systematic procedure at one-loop, that relates sesqui-linear torus partition functions to linear annulus partition functions 
\begin{equation}
{\cal T} = \sum_{h, \bar{h}} N_{h, \bar{h}} \chi_h(\tau) \bar\chi_{\bar{h}}(\bar\tau) \rightarrow
{\cal A}_{a,b} = \sum_{h} A^{h}_{a,b} \chi_h(it) 
\end{equation}
with $a,b$ labelling the boundary conditions (D-branes) and $A^{h}_{a,b}$ counting the number of ways the primary field of dimension $h$ appears in the spectrum of open strings joining $a$ with $b$, one can write an ansatz for the disk correlator
\begin{equation}
{\cal G}^\textup{disk}_{a,b,c,d}(x) =  {\rm Tr}(T_{a,b} T_{b,c}T_{c,d} T_{d,a}) \sum_{h_o} C^{h_o}_{a,c} {\cal F}_{h_o}(x) 
\end{equation}
In principle the set $h_o$ does not need to coincide with the set $h$, due to Wilson lines, shifts and angles, yet the enumeration follows the one in the closed string case.

Unitarity and planar duality put severe constraints. The crucial point is that the coefficient $C^{h_o}_{ab|cd}$ be chosen in such a way that the correct open string spectrum be exchanged not only in the $s$ channel, that is exposed by the limit $x\rightarrow 0$ (open strings from $a$ to $c$), but also in the $t$ channel which is not manifest in the above parametrisation, since it requires $x\rightarrow 1$ (open strings from $b$ to $d$). The problem has been solved in non-trivial contexts for minimal models \cite{Bianchi:1991rd} and later on for WZW models \cite{Pradisi:1995pp}. In the case at hand, \ie correlator of (excited) twist fields, one can use an alternate route, known as the `doubling' trick, and arrive at consistent expressions for the open-string correlators \cite{Cvetic:2003ch,Erler:1992gt,Anastasopoulos:2013sta}. 
The detailed description of our strategy and the computations of the relevant correlators can be found in the appendix \ref{sect:twist_fields}.
 
Before proceed to compute scattering amplitudes and extract gauge and Yukawa couplings, generalising previous work on the subject \cite{Cvetic:2003ch, Cremades:2003qj, Abel:2003vv, Lust:2004cx, Cvetic:2007ku}, we would like to spell out the possible choices of angles for the configurations under attention. Although we have in mind at least two stacks of branes intersecting with one small angle, our analysis will be rather general.

\section{Scattering amplitudes} \label{sect:scattering_amplitudes}

Given the knowledge of the BRST invariant vertex operators for both massless and massive (but light) string states one can proceed and compute scattering amplitudes and extract gauge and Yukawa couplings. In the computation of Yukawa couplings, we rely on the  factorization of  scattering amplitudes that expose gauge boson exchange and that can be correctly normalised, relying on the field theory limit. Choosing the amplitudes appropriately, one can fix the normalizations of both massless and massive charged vertex operators that involve twist-fields. The desired Yukawa's can be derived via factorization of the same amplitudes in different channels or from altogether different amplitudes without external or intermediate gauge bosons. As we will see, in the latter case, the knowledge of certain Yukawa's, previously computed, replaces the knowledge of the gauge couplings in fixing unambiguously the normalizations.

Vertex operators for gauge bosons and gaugini carry Chan-Paton factors for the Adjoint of the gauge group $U(N)$ \footnote{We will not consider $SO$ or $Sp$ groups in this paper.} corresponding to open strings with both ends on the same stack of D-branes. Open strings with ends on different stacks transform in the bi-fundamental (${\bf N_a, \bar{N}_b}$). Due to total anti-symmetry of the tri-linear gauge vector couplings, the relevant Chan-Paton factor turns out to be  
\begin{equation}
{\rm Tr} (T^A [T^B, T^C]) = i f^{ABC}
\end{equation}

For coupling of a gauge boson of $U(N_a)$  to matter fields in the  (${\bf N_a, \bar{N}_b}$), the relevant C-P factor is
\begin{equation}
{T_{(a)}^A}^{i_a}{}_{j_a} \delta_{i_b}{}^{j_b}
\end{equation}
For color-ordered 4-pt functions the Chan-Paton factor is in general
\begin{equation}
\sum_{i_a,j_b,k_c,l_d} {T_{(a,b)}}^{i_a}{}_{j_b} {T_{(b,c)}}^{j_b}{}_{k_c} {T_{(c,d)}}^{k_c}_{l_d} {T_{(d,a)}}^{l_d}_{i_a} = {\rm Tr}(T_{(a,b)} T_{(b,c)} T_{(c,d)} T_{(d,a)})
\end{equation}
for one and two independent angles the situation simplifies to
\begin{equation}
\sum_{i_a,j_b,k_a,l_b} {T_{(a,b)}}^{i_a}{}_{j_b} {T_{(b,a)}}^{j_b}{}_{k_a} {T_{(a,b)}}^{k_a}_{l_b} {T_{(b,a)}}^{l_b}_{i_a} = {\rm Tr}(T_{(a,b)} T_{(b,a)} T_{(a,b)} T_{(b,a)})
\end{equation}
or to 
\begin{equation}
\sum_{i_a,j_b,k_a,l_c} {T_{(a,b)}}^{i_a}{}_{j_b} {T_{(b,a)}}^{j_b}{}_{k_a} {T_{(a,c)}}^{k_a}_{l_c} {T_{(c,a)}}^{l_c}_{i_a} = {\rm Tr}(T_{(a,b)} T_{(b,a)} T_{(a,c)} T_{(c,a)})
\end{equation}
Henceforth we will not explicitly write the above Chan-Paton factors in order not to burden formulae too much.

\subsection{Yang-Mills couplings} \label{sect:YM_couplings}
The computation of amplitudes with external gauge bosons allows us not only determine the Yang-Mills couplings $\gYM{a}$, in terms of the string coupling $g_\textup{op}$ and the geometric data of the brane configuration, but also the normalizations of the vertex operators, denoted by $C_{\c_0^{ab}}$, $C_{\f_0^{ab }}$ and $C_{A_a}$ henceforth. For simplicity we consider color-ordered amplitudes and drop Chan-Paton factors, as already announced. The amplitude $\calA(A_a,A_a,A_a)$ vanishes on-shell but we can relax the kinematic conditions, for instance taking complex momenta, to have a non-zero result. Fixing $\calA(A_a,A_a,A_a)$ to be $\gYM{a}/\sqrt{2}$ up to kinematic factors, we extract the relation between the normalization of the vector boson vertex $C_{A_a}$ and the normalisation of the disk $C_{D^2_a}\sim g_{\rm op}^{-2} = g_{\rm cl}^{-1}$. The amplitude $\calA(A_a,A_a,A_a,A_a)$ can be used to fix the relation between $C_{A_a}$ and $\gYM{a}$, after factorizing it either in the $s$- or in the $t$-channel. The relations that one finds read
\begin{equation}
\label{eq:g_YM_many_bosons}
\gYM{a}= \frac{C_{A_a}}{\sqrt{2\ap}}\, g_\textup{op} \qquad 2 \ap C_{D^2_a} C_{A_i}^2=1
\end{equation}
In order to have a real gauge coupling constant, the vertex operator normalization $C_{A_a}$ must be real as well as $C_{D^2_a}$. The 3-pt amplitudes with two massless fermions or scalar bosons and one gauge boson then give us the relation between vertex operators normalizations $C_{\c^{ab}_0}$ and $C_{\f^{ab}_0}$ in terms of $C_{A_a}$. Imposing that the color ordered amplitudes $\calA(\f_0^{ab},\bar{\f}_0^{ba},A_a)$ and $\calA(\c_0^{ab},\cB_0^{ba},A_a)$ be $\gYM{a}/\sqrt{2}$ up to kinematic factors, where $\cB_0^{ba}$ is the charge conjugate of $\cB_0^{ab}$, one finds
\begin{equation}
|C_{\c^{ab}_0}|^2 = \sqrt{\ap} C_{A_a} C_{A_b} \qquad,\qquad |C_{\f^{ab}_0}|^2 = C_{A_a} C_{A_b}
\end{equation}
SUSY transformations relate the vertex operators $V_{\f_0^{ab}}$ and $V_{\c_0^{ab}}$, thus we can identify the phases of the normalization constants $C_{\f^{ab}_0}$ and $C_{\c^{ab}_0}$.

\subsubsection{The amplitude \texorpdfstring{$\calA(\bar{\y}_0,\y_0,\bar{\y}_0,\y_0)$}{A(psi0,psi0,psi0,psi0)} [One Angle]}
The amplitude $\calA(\bar{\y}_0,\y_0,\bar{\y}_0,\y_0)$ depends on only one independent angle between branes $b$ and $c$ and allows us to fix $C_{A_b}$, as well as $C_{\c_0^{bc}}$, $C_{\f_0^{bc}}$ and $\gYM{b}$. Denoting the fermion $\c^{bc}_0 =\y_0$ and its charge conjugate as $\bar{\c}^{cb}_0=\bar{\y}_0$ and assuming that they are all localised at the same intersection point \ie $f_1=f_2=f_3=f_4=f_{\y}$, there are no shifts in the classical action. As a result the lightest state in both channels is a gauge boson, either $A_b$ or $A_c$. This allows to fix the normalization of the correlator of four un-excited twist fields $\s$ with one independent angle. This normalization appears in all the 4-pt twist field correlators\footnote{On-shell fermion polarisations such as $\y_0, \bar{\y}_0$ have mass dimension 1 in $D=4$.}
\begin{figure} [t]
\centering
\begin{tikzpicture}
\draw [red] (-0.2,0) -- (2.2,0) (2-0.2,2) -- (4.2,2);
\draw [blue] (45:{-0.2*1.4}) -- (2,2)--++(45:{0.2*1.4}) (2,0)++(45:{-0.2*1.4}) -- (4,2)--++(45:{0.2*1.4});
\fill (0,0) circle (2 pt) (4,2) circle (2 pt) (2,0) circle (2 pt) (2,2) circle (2 pt);
\draw (0+15 pt,0) arc (0:45:15 pt);
\node at (23:25 pt) {$\a$};
\node at (1,0-7 pt) {$c$};
\node at (3,7 pt+2 cm) {$c$};
\node at (1 cm-10 pt,1) {$b$};
\node at (3 cm+10 pt,1) {$b$};
\node at (-20 pt,-10 pt) {$\y_0(2)$};
\node at (2 cm+10 pt,-10 pt) {$\bar{\y}_0(3)$};
\node at (4 cm+20 pt,2 cm+10 pt) {$\y_0(4)$};
\node at (2 cm-10 pt,2 cm+10 pt) {$\bar{\y}_0(1)$};
\end{tikzpicture}
\caption{A configuration of branes and intersecting points describing the scattering between $\y_0$ and $\bar{\y}_0$.}
\label{fig:one_angle}
\end{figure}
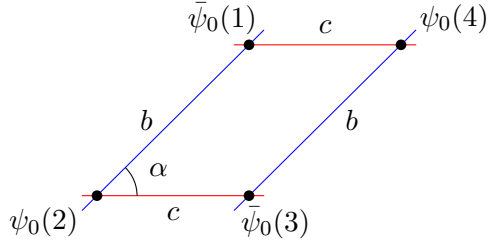
\begin{equation}
\begin{split}
\calA(\bar{\y}_0,\y_0,&\bar{\y}_0,\y_0)=|C_{\y_0}|^4 \sqrt{C_{D^2_b} C_{D^2_c}} g_\textup{op}^2 \y_0(2) {\cdot}\y_0(4) \yB_0(1) {\cdot} \yB_0(3) \int_0^1 dx\,  x^{\ap s-1}(1-x)^{\ap t-1} \times \\
& \times \prod_{I=1}^3\dfrac{4\p^2 \ap K_I^{c,b}}{L_{b,I} L_{c,I} F_{\a_I}(x)} \sum_{n_I,m_I} \exp\lf[-\frac{\p t_I(x)}{\sin \p |a^I_{bc}|} \lf(\frac{4\p^2 \ap}{ L_{c,I}^2} m_I^2+\frac{\sin^2 \p |a^I_{bc}|}{4\p^2\ap} n_I^2 L_{b,I}^2\rg) \rg]
\end{split}
\end{equation}
where the Mandelstam variables are $s=-(k_1+k_2)^2$ and $t=-(k_1+k_4)^2$ and $K_I^{c,b}$ denotes the normalization of the four twist-fields correlator with one-independent angle. $F_{\a_I}(x)$ are the hypergeomtric functions\footnote{We recall that in the present setting $a^{1,2}_{bc}>0$ while $a^3_{bc}<0$.}
\begin{equation}
F_{\a_I}(x)=\Fab{\a_I}{1{-}\a_I}{1}{x} \qquad \text{with} \quad \a_{1,2}=a_{bc}^{1,2} \,\, , \,\,\a_3=1+a^3_{bc}=1-|a^3_{bc}|
\end{equation}
The (massless) poles in the $s$ and $t$ channels are exposed by the limits $x\to 0,1$. In the $s$ channel ($x\to 0$), the leading term  obtains by minimizing the exponential thus taking $n_I=m_I=0$
\begin{equation}
\calA(\bar{\y}_0,\y_0,\bar{\y}_0,\y_0)\xrightarrow{s\to 0} \dfrac{\gYM{c}^2}{2} \, \y_0(2)\s_\m \bar{\y}_0(1)\, \dfrac{1}{s}\, \y_0(4) \s^\m \bar{\y}_0(3) +\dots
\end{equation}
one obtains the following expression for $\gYM{c}$
\begin{equation}
\gYM{c} =\dfrac{|C_{\y_0}|^2}{\sqrt{\ap}} (C_{D^2_b} C_{D^2_c})^{1/4} g_\textup{op} \prod_{I=1}^3 \sqrt{\dfrac{ 4\p^2\ap K_I^{c,b}}{L_{b,I} L_{c,I}}}
\end{equation}
In order to  correctly perform the limit $x\to 1$ and extract $\gYM{b}$ in the $t$ channel, it is convenient to Poisson resum over both indices $m_I$ and $n_I$. Exactly as in the $s$ channel, one can compute the leading term and factorize it on 3-pt amplitudes thus obtaining
\begin{equation}
\gYM{b} = \frac{|C_{\y_0}|^2}{\sqrt{\ap}} (C_{D^2_b} C_{D^2_c})^{1/4} g_\textup{op} \prod_{I=1}^3 \sqrt{\dfrac{4\p^2 \ap K_I^{c,b}}{L_{b,I}^2}}
\end{equation}
The ratio between $\gYM{c}$ and $\gYM{b}$ depends only by the branes' lengths hence the product $C^2_{A_a} \prod_I L_{a,I}$ is brane-independent. This combined with some dimensional analysis allows us to fix the gauge couplings and $C_{A_a}$ to be
\begin{gather}
\gYM{a}=g_\textup{op} \prod_I \sqrt{\dfrac{\sqrt{\ap}}{ L_{a,I}}}  \qquad C_{A_a}=\sqrt{2\ap}\prod_I \sqrt{\frac{\sqrt{\ap}}{L_{a,I}}}\\
C_{\c^{bc}_0}= e^{i \g^{bc}_0} (\ap)^{1/4} \sqrt{2\ap} \prod_I \lf[\frac{\ap}{L_{b,I}L_{c,I}} \rg]^{1/4}  \qquad
C_{\f^{bc}_0}= e^{i \g^{bc}_0} \sqrt{2\ap} \prod_I \lf[\frac{\ap}{L_{b,I}L_{c,I}} \rg]^{1/4} 
\end{gather}
where $\g^{bc}_0$ is an unfixed phase. 
As a result the normalization of $K_I^{c,b}$ turns out to be
\begin{equation}
\label{eq:K_normalization}
K_I^{c,b}=\frac{\sqrt{L_{b,I} L_{c,I}} L_{b,I}}{4\p^2 \ap}
\end{equation}
This and other expressions later on look slightly asymmetric in the `indices' $a,b,c,d$. This is due to the privileged role played by the $s$-channel ($x\rightarrow 0$) but the physical content is totally symmetric. 

\subsection{Yukawa couplings for massless external legs} \label{sect:yukawa_massless_states}
In this section we will review the computation of the amplitudes with massless external legs that produce informations on the Yukawa couplings.

\subsubsection{The amplitude \texorpdfstring{$\calA(\y_0,\c_0,\f_0)$}{A(psi0,chi0,phi0)}}

The selection rules associated to the conservation of the $U(1)$ R-charges and coded in the correlators of the exponentials $e^{i q_I \vf_I}$ coincide with the conditions on the intersection angles \eqref{eq:non_vanish_Yukawas_consitions} given in section \ref{sect:setupandVOs} for the amplitude $\calA(\f_0,\y_0,\c_0)$ not to vanish.
\begin{figure} [t]
\centering
\begin{tikzpicture}
%\draw [step=0.25] (0,0) grid (8,4);
\fill [yellow] (0,0) -- (2,0) --(2,2);
\draw [dashed,gray] (0,-0.2) -- (0,4.2) (8,-0.2) -- (8,4.2);
\draw [red] (-0.2,0) -- (8.2,0) (-0.2,4) -- (8.2,4);
\draw [blue] (45:{-0.2*1.4}) -- (4,4)--++(45:{0.2*1.4}) 
(8,0)++(45:{-0.2*1.4}) -- (8,0)--++(45:{0.2*1.4})
(4,0)++(45:{-0.2*1.4}) -- (8,4)--++(45:{0.2*1.4}) 
(0,4)++(45:{-0.2*1.4}) -- (0,4)--++(45:{0.2*1.4});
\draw [green] (2,-0.2) -- (2,4.2);
\fill (0,0) circle (2 pt) (8,0) circle (2 pt) 
(0,4) circle (2 pt) (8,4) circle (2 pt) ;
\fill (-2 pt+4 cm,-2 pt) rectangle (2 pt+4 cm,2 pt) (-2 pt+4 cm,-2 pt+4 cm) rectangle (2 pt+4 cm,2 pt+4cm);
\draw (-2pt+2 cm,-2pt) rectangle (2pt+2 cm,2pt)
(-2pt+2cm,4cm-2pt) rectangle (2pt+2cm,4cm+2pt);
\draw (2,2) circle (2pt);
\node at (-10 pt,-15 pt) {$\y_0(1)$};
\node at (2 cm,-15 pt) {$\c_0(2)$};
\node at (2 cm +20 pt,2 cm) {$\f_0(3)$};
\end{tikzpicture}
\caption{The minimal area defined by the three intersection points $f_{\y}$, $f_{\c}$ and $f_{\f}$}
\label{fig:yukawa}
\end{figure}
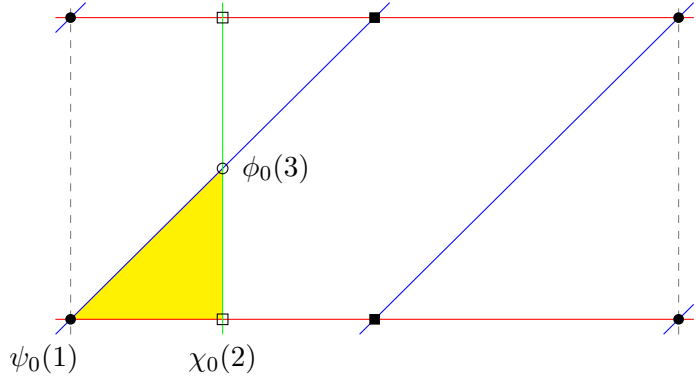
The 3-pt amplitude reads
\begin{equation}
\begin{split}
\calA(\y_0^{bc},\c_0^{ca},\f_0^{ab})&= \frac{1}{\sqrt{2}} g_\textup{op} \f_0(3)\, \y_0(1) {\cdot} \c_0(2) \prod_{I=1}^3 \frac{N_I^{abc} \ap^{1/4}}{(L_{a,I} L_{b,I} L_{c,I})^{1/6}} \times \\
& \times \sum_{n_I} \exp\lf[-\frac{1}{2\p \ap}\frac{\sin \p |a^I_{bc}| \sin \p |a^I_{ca}|}{2\sin \p |a^I_{ab}|}(f_{\c \y,I}+n_I \tilde{L}_{c,I})^2 \rg]
\end{split}
\end{equation}
where $f_{\c \y,I}=f_{\c,I}-f_{\y,I}$. Since the contribution of the third torus to the classical action involves a factor $\sin \p (1-a^3_{ab})$, one must restrict the angle to be in the range $-1<a^3_{ca}<0$. Neglecting wrapping states, $n_I=0$, and dropping the polarizations we obtain the Yukawa coupling $Y_{000}$ between $\f_0$, $\y_0$ and $\c_0$:
\begin{equation}
\label{eq:yukawa_def}
Y_{000}=e^{i(\g_0^{ab}+\g_0^{bc}+\g_0^{ca})}\frac{g_\textup{op}}{\sqrt{2}} \prod_{I=1}^3 \frac{N_I^{abc} \ap^{1/4}}{(L_{a,I} L_{b,I} L_{c,I})^{1/6}} \exp\lf[-\frac{A^{(I)}_{\f \y \c}}{2\p\ap}\rg]
\end{equation}
where $A^{(I)}_{\f \y \c}$ is the area of the triangle defined by the three points $f_{\y,I}$, $f_{\c,I}$ and $f_{\f,I}$ in the torus $I$ given by
\begin{equation}
A^{(I)}_{\f \y \c}= \frac{\sin \p |a^I_{bc}| \sin \p |a^I_{ca}|}{2\sin \p |a^I_{ab}|} f_{\c \y,I}^2
\end{equation}

\subsubsection{The amplitude \texorpdfstring{$\calA(\bar{\y}_0,\y_0,\c_0,\bar{\c}_0)$}{A(psi0,psi0,chi0,chi0)} [Two Angles]}

The computation of the amplitude $\calA(\bar{\y}_0,\y_0,\c_0,\bar{\c}_0)$ with two-independent angles yields
\begin{equation}
\begin{split}
\calA(\bar{\y}_0^{cb},\y_0^{bc},\c_0^{ca},\bar{\c}_0^{ac})&= g_\textup{op}^2 \ap \y_0(2) {\cdot}\c_0(3) \bar{\y}_0(1) {\cdot} \bar{\c}_0(4) \int_0^1 dx\, x^{\ap s-1}(1-x)^{\ap t-1} \times \\
& \times \prod_{I=1}^3 \frac{4\p^2 K_I^{c,ab} \ap}{L_{a,I}^{1/4} L_{b,I}^{5/4} L_{c,I}^{1/2}}
\frac{\sqrt{\ap}}{L_{c,I} G_1^{(I)} (x)}
 \sum_{n_I,m_I} e^{-S_\textup{Ham}^{(I)}(m_I,n_I)}
\end{split}
\end{equation}
where $G_1^{(I)} (x)$ are hypergeometric functions depending on the angles defined in the appendix \eqref{eq:def_functions_G}, $K_I^{c,ab}$ is an overall normalization constant and $S_\textup{Ham}^{(I)}(m_I,n_I)$ is the classical action in  `hamiltonian' form:
\begin{equation}
S_\textup{Ham}^{(I)}(m_I,n_I)=
\frac{\p t_I(x)}{\sin \p |a^I_{bc}|} \lf(\frac{4\p^2 \ap m^2_I}{L_{c,I}^2} +\frac{ \sin^2 \p |a^I_{bc}| L_{b,I}^2}{4\p^2 \ap} \frac{I_{ca,I}^2 n_I^2}{\gcd^2(|I_{bc,I}|,|I_{ca,I}|)} \rg)-2\p i \frac{m_I}{L_{c,I}} f_{\c \y,I}
\end{equation}
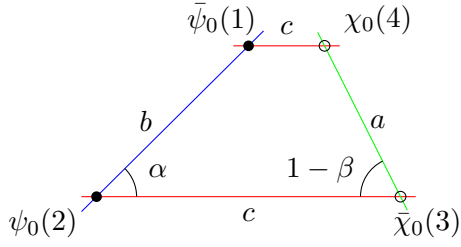
\begin{figure} [t]
\centering
\begin{tikzpicture}
\draw [red] (-0.2,0) -- (4.2,0) (2-0.2,2) -- (3.2,2);
\draw [blue] (45:{-0.2*1.4}) -- (2,2)--++(45:{0.2*1.4});
\draw [green] (4,0)++({atan2(-1,2)}:-0.2) -- (3,2)--++({atan2(-1,2)}:0.2);
\fill (0,0) circle (2 pt) (2,2) circle (2 pt);
\draw (3,2) circle (2pt) (4,0) circle (2pt);
\draw (0+15 pt,0) arc (0:45:15 pt);
\draw (4cm-15 pt,0) arc (180:{atan2(-1,2)}:15 pt);
\node at (23:25 pt) {$\a$};
\node at (4cm-30 pt,10 pt) {$1-\b$};
\node at (2,0-7 pt) {$c$};
\node at (2.5,7 pt+2 cm) {$c$};
\node at (1 cm-10 pt,1) {$b$};
\node at (3 cm+20 pt,1) {$a$};
\node at (-20 pt,-10 pt) {$\y_0(2)$};
\node at (4 cm+10 pt,-10 pt) {$\bar{\c}_0(3)$};
\node at (3 cm+20 pt,2 cm+10 pt) {$\c_0(4)$};
\node at (2 cm-10 pt,2 cm+10 pt) {$\bar{\y}_0(1)$};
\end{tikzpicture}
\caption{A configuration of branes describing the scattering between the spinors $\y_0$ and $\c_0$.}
\label{fig:two_angle}
\end{figure}
In the $s$ channel the amplitude shows a gauge boson propagating between parallel branes of type $c$, we can use the factorization in this channel to fix $K_I^{c,ab}$:
\begin{equation}
K_I^{c,ab}=\frac{L_a^{1/4}L_b^{5/4} L_{c,I}^{1/2}}{(2\p)^2 \ap}
\end{equation}
The factorization in the $t$ channel exhibits Yukawa couplings, thus it is convenient to have the classical action in the form of an area divided by $\ap$. To this end we perform a Poisson resummation over $m_I$. 
\begin{equation}
\label{eq:By_y_c_Bc_lagrangian_form}
\begin{split}
\calA(\bar{\y}_0,\y_0,\c_0,\bar{\c}_0) &= g_\textup{op}^2 \ap \y_0(2) {\cdot}\c_0(3) \bar{\y}_0(1) {\cdot} \bar{\c}_0(4) \int_0^1 dx\, x^{\ap s-1}(1-x)^{\ap t-3/2} \times \\
& \times \prod_{I=1}^3 \sum_{\tilde{m}_I,n_I} \dfrac{e^{-S_\textup{Lagr}^{(I)}(\tilde{m}_I,n_I)}}{2\p \sqrt{I_I(x)}}
\end{split}
\end{equation}
where $I_I(x)$ are combinations of hypergeometric functions defined in the appendix \eqref{eq:def_function_I} and $S_\textup{Lagr}^{(I)}(\tilde{m}_I,n_I)$ is the classical action in  `lagrangian' form
\begin{equation}
S_\textup{Lagr}^{(I)}(\tilde{m}_I,n_I)= \frac{\sin \p |a^I_{bc}|}{4 \p \ap}\lf[\frac{1}{t_I(x)}  (\tilde{m}_I L_{c,I}+f_{\c \y,I} )^2+t_I(x) \lf(\dfrac{I_{ca,I}}{\gcd(|I_{bc,I}|,|I_{ca,I}|)} L_{b,I}  n_I \rg)^2 \rg]
\end{equation}
The limit $x\to 1$ has two kinds of contributions: one purely quantum, due to the expansion of $I_I(x)$, and one classical, due to the expansion of $t(x)$ in the action. The first three orders correspond to factorizations on the poles for the states $\f_0$, $\f_1$ and $\f_2$. Imposing  
\begin{equation}
\calA(\bar{\y}_0,\y_0,\c_0,\bar{\c}_0)\xrightarrow{t\to n\, a^1_{ab}/\ap} |Y_{n00}|^2 \y_0(2) {\cdot}\c_0(3) \dfrac{1}{t- n\,  a^1_{ab}/\ap} \bar{\c}_0(4){\cdot} \bar{\y}_0(1) 
\end{equation}
The relative importance of the various terms in the expansion depends the angles $a^1_{ab}$, $a^2_{ab}$ and $1+a^3_{ab}$. We choose an ordering of the magnitudes of the angles in which the angles in the first torus are smaller.
\begin{equation}
a_{ab}^1<2 a_{ab}^1 <a_{ab}^2<1+a_{ab}^3=1-|a_{ab}^3|
\end{equation}
Combining these conditions with $a^1_{ab}+a^2_{ab}+a^3_{ab}=0$ we obtain two equivalent equations
\begin{equation}
3 a^1_{ab}<|a^3_{ab}|<\frac{3}{2} a^2_{ab} \qquad 2|a^3_{ab}|-1< a^1_{ab} < 1-2 a^2_{ab}
\end{equation}
The Yukawa's extracted by factorization are given by
\begin{align}
|Y_{000}|&= g_\textup{op} (2\p)^{-3/4} [\G_{1-a^1_{ab},1-a^1_{bc},-a^1_{ca}} \G_{1-a^2_{ab},1-a^2_{bc},-a^2_{ca}} \G_{-a^3_{ab},-a^3_{bc},-a^3_{ca}}]^{1/4} \times \nn \\
&\quad \times \prod_{I=1}^3 \exp\lf[- \frac{ A_{\f \y \c}^{(I)} }{ 2 \p \ap} \rg] \\
|Y_{100}|&=\frac{ |Y_{000}|}{\sqrt{a^1_{ab}}} [\G_{1-a^1_{ab},1-a^1_{bc},-a^1_{ca}}]^{1/2} \sqrt{\frac{2  A_{\f \y \c}^{(1)} }{\p\ap}}  \\
|Y_{200}|&= \frac{|Y_{000}|}{\sqrt{2} a^1_{ab}}\G_{1-a^1_{ab},1-a^1_{bc},-a^1_{ca}} \lf|\frac{2 A_{\f \y \c}^{(1)} }{\p\ap}-1\rg|
\end{align}
where
\begin{equation}
\G_{a,b,c}=\dfrac{\G(a)\G(b)\G(c)}{\G(1-a)\G(1-b)\G(1-c)}
\end{equation}
Comparing Yukawa $Y_{000}$ with the previous result \eqref{eq:yukawa_def} we obtain the normalization of the 3-pt twist fields correlator
\begin{equation}
|N_I^{abc}|= 2^{1/6} \lf(\frac{\G_{1-\a_I,\b_I,1+\a_I-\b_I}}{2\p} \rg)^{1/4} \lf[\frac{L_{a,I}}{\sqrt{\ap}}\frac{L_{b,I}}{\sqrt{\ap}}\frac{L_{c,I}}{\sqrt{\ap}}\rg]^{1/6}
\end{equation}
We cannot determine the phases of the Yukawa couplings $Y_{n00}$, but we have a relation among the phase of $Y_{000}$, the normalizations $N_I^{abc}$ and the phases of the vertex operators
\begin{equation}
\arg Y_{000} = \g_0^{ab}+\g_0^{bc}+\g_0^{ca}+\sum_I \arg N^{abc}_I
\end{equation}

\subsubsection{The amplitudes \texorpdfstring{$\calA(\bar{\f}_0,\f_0,\y_0,\bar{\y}_0)$}{A(phi0,phi0,psi0,psi0)} and \texorpdfstring{$\calA(\bar{\c}_0,\c_0,\f_0,\bar{\f}_0)$}{A(chi0,chi0,phi0,phi0)}}

There are two 4-pt amplitudes that factorise on the Yukawas $Y_{010}$, $Y_{001}$, $Y_{020}$ and $Y_{002}$, they are $\calA(\bar{\c}_0,\c_0,\f_0,\bar{\f}_0)$ and $\calA(\bar{\f}_0,\f_0,\y_0,\bar{\y}_0)$. Computing these amplitudes require the knowledge of $V^{(0)}_{\f_0}$, that should be determined including its precise normalization. However there is a simpler way that relies on the Ward identities associated to supersymmetry transformations. Indeed we may compute amplitudes with four fermions instead of amplitudes with two fermions and two bosons. As shown in \cite{Anastasopoulos:2016cmg} the states $\{\f^{ij}_0,\c^{ij}_0\}$ form an $\calN=1$ multiplet.
\begin{gather}
\label{eq:SUSY_vertex_fermions}
[\e{\cdot}Q,V_{\c^{ij}_0 (k)}]= V_{\d_\e \f^{ij}_0 = \e {\cdot} \c^{ij}_0(k)} \qquad 
[\e{\cdot}Q,V_{\bar{\c}^{ji}_0(k)}]= 0 \\
\label{eq:SUSY_vertex_bosons}
[\e{\cdot}Q,V_{\f^{ij}_0(k)}]= 0 \qquad
[\e{\cdot}Q,V_{\bar{\f}^{ji}_0(k)}]= V_{\d_\e \cB^{ji}_0=\bar{\s}^\m \e k_\m \bar{\f}^{ji}_0(k)}
\end{gather}
In order to obtain a relation between $\calA(\bar{\f}^{ba}_0,\f^{ab}_0,\c^{bc}_0,\bar{\c}^{cb}_0)$ and $\calA(\bar{\c}^{ba}_0,\c^{ab}_0,\c^{bc}_0,\bar{\c}^{cb}_0)$ we consider the commutator
\begin{equation}
[\e{\cdot}Q, V_{\bar{\f}^{ba}_0}\, V_{\c^{ab}_0}\, V_{\c^{bc}_0}\, V_{\bar{\c}^{cb}_0}]=0
\end{equation}
Choosing $\e=\c^{bc}_0(3)$ one get the Ward identity
\begin{equation}
\calA(\bar{\f}^{ba}_0,\f^{ab}_0,\c^{bc}_0,\bar{\c}^{cb}_0)=\frac{\c^{ab}_0(1){\cdot}\c^{bc}_0(3)}{\c^{bc}_0(3){\cdot} \c^{ab}_0(2)} \calA(\bar{\c}^{ba}_0,\c^{ab}_0,\c^{bc}_0,\bar{\c}^{cb}_0)
\end{equation}
Similarly we find an identity for $\calA(\bar{\c}_0,\c_0,\f_0,\bar{\f}_0)$ with the choice $\e=\c^{ca}_0(2)$
\begin{equation}
\calA(\bar{\c}^{ac}_0,\c^{ca}_0,\f^{ab}_0,\bar{\f}^{ba}_0)= 
\frac{\c^{ca}_0(2){\cdot}\c^{ab}_0(4)}{\c^{ab}_0(3){\cdot}\c^{ca}_0(2)}
\calA(\bar{\c}^{ac}_0,\c^{ca}_0,\c^{ab}_0,\bar{\c}^{ba}_0)
\end{equation}
The computation of the amplitude with four fermions is completely equivalent to the previous case thus we only display the amplitudes after the Ward identities
\bea
\calA(\bar{\c}_0,\c_0,\f_0,\bar{\f}_0) &=& \ap g_\textup{op}^2 \f_0(3) \bar{\f}_0(4) \frac{k_3^\m -k_4^\m }{2}\c_0(2) \s_\m \cB_0(1) \\
&&~~~~~~~~~~~~~~~~~ \times \int_0^1 dx\, x^{\ap s-1}(1-x)^{\ap t-3/2} \prod_{I=1}^3 \sum_{\tilde{m}_I,n_I} \dfrac{e^{-S_\textup{Lagr}^{(I)}(\tilde{m}_I,n_I)}}{2\p \sqrt{I_I(x)}} \nn\\
\calA(\bar{\f}_0,\f_0,\y_0,\bar{\y}_0) &=& \ap g_\textup{op}^2 \f_0(2) \bar{\f}_0(1)\frac{k_2^\m -k_1^\m }{2} \yB_0(4) \s_\m \y_0(3) \\
&&~~~~~~~~~~~~~~~~~ \times  \int_0^1 dx\, x^{\ap s-1}(1-x)^{\ap t-3/2} \prod_{I=1}^3 \sum_{\tilde{m}_I,n_I} \dfrac{e^{-S_\textup{Lagr}^{(I)}(\tilde{m}_I,n_I)}}{2\p \sqrt{I_I(x)}}\nn
\eea
The expansion of the first amplitudes is governed by the relative magnitudes of the angles $a_{bc}^1$, $a_{bc}^2$ and $1+a_{bc}^3$ while the second by $1+a_{ca}^1$, $1+a_{ca}^2$ and $1+a_{ca}^3$. We impose the following relations for the $bc$ and $ca$ angles
\begin{equation}
a_{bc}^1<2 a_{bc}^1 <a_{bc}^2<1+a_{bc}^3 \qquad |a^1_{ca}|<2|a^1_{ca}|<|a^2_{ca}|<|a^3_{ca}|
\end{equation}
Exactly as for the angles $ab$ in the previous amplitude we can find alternative relations for
\begin{gather}
3 a^1_{bc}<|a^3_{bc}|<\frac{3}{2} a^2_{bc} \qquad 2|a^3_{bc}|-1< a^1_{bc} < 1-2 a^2_{bc} \\
3|a^1_{ca}|<2-|a^3_{ca}|<\frac{3}{2} |a^2_{ca}| \qquad 2|a^2_{ca}| <2-|a^1_{ca}| < 2 |a^3_{ca}|
\end{gather}
We note that the smaller angles for the intersections $ca$ is $1+a^3_{ca}$. The factorization of these amplitudes in the $t$ channel yields the desired Yukawa's
\begin{align}
|Y_{010}|&= \frac{|Y_{000}|}{\sqrt{a^1_{bc}}}[\G_{1-a^1_{ab},1-a^1_{bc},-a^1_{ca}}]^{1/2} \sqrt{\frac{2  A_{\f \y \c}^{(1)} }{\p\ap}}  \\
|Y_{001}|&= \frac{|Y_{000}|}{\sqrt{1+a^3_{ca}}} [\G_{-a^3_{ab},-a^3_{bc},-a^3_{ca}}]^{1/2} \sqrt{\frac{2  A_{\f \y \c}^{(3)} }{\p\ap}} \\
|Y_{020}|&= \frac{|Y_{000}|}{\sqrt{2} a^1_{bc}}\G_{1-a^1_{ab},1-a^1_{bc},-a^1_{ca}}\lf|\frac{2  A_{\f \y \c}^{(1)} }{\p\ap}-1\rg|  \\
|Y_{002}|&= \frac{|Y_{000}|}{\sqrt{2} (1+a^3_{ca})}\G_{-a^3_{ab},-a^3_{bc},-a^3_{ca}} \lf|\frac{2  A_{\f \y \c}^{(3)} }{\p \ap}-1\rg| 
\end{align}

\subsection{Yukawa couplings from amplitudes with massive external legs} \label{sect:yukawa_massive_states}
The factorization of 4-pt amplitudes with massive external legs yields other Yukawa's that involve more than one massive particle. As done previously we can fix the normalizations of the vertex operators from amplitudes that include gauge bosons. First we note that an excited fermionic state cannot decay in a un-excited state and a gauge bosons
\begin{equation}
\label{eq:chi1_BARchi0_A_vanishes}
\calA(\c_1^{ab},\cB_0^{ba},A_a)=0 \qquad \calA(\cB_1^{ba},\c_0^{ab},A_b)=0
\end{equation}
This is due to conformal invariance that forbids a non-vanish 2-pt correlator between the twist fields $\s_\a$ and $\t_\b$. The amplitudes $\calA(\c_1^{i,j},\cB_1^{j,i},A_i)$ and $\calA(\bar{\tilde{\c}}_1^{\dagger\,i,j},\tilde{\c}_1^{\dagger\,j,i},A_i)$ must reproduce an interaction between a gauge field and a Dirac spinor. This gives rise to the two conditions
\begin{gather}
|C_{\c_1^{ab}}|^2 = |C_{\tilde{\c}_1^{ab}}|^2= \sqrt{\ap} C_{A_a} C_{A_b}(=|C_{\c_0^{ab}}|^2)
\end{gather}
thus the normalization of the vertex operator for an excited spinor is the same as for an un-excited spinor.

\subsubsection{The amplitudes \texorpdfstring{$\calA(\y_0,\c_0,\f_1)$}{A(psi0,chi0,phi0)} and \texorpdfstring{$\calA(\cB_0,\yB_0,\bar{\f}_1)$}{A(psi0,chi0,phi0)}}
We already know the Yukawa coupling $|Y_{100}|$ but the amplitudes $\calA(\y_0,\c_0,\f_1)$ and its conjugate correspondent $\calA(\cB_0,\yB_0,\bar{\f}_1)$ help us to fix the normalization of the correlator between one excited twist field and two un-excited ones. In turn the normalization depends on the OPE coefficient $C_{\der Z,\s}^\t(\a)$, for more details \eqref{eq:OPE_der_Z_and_sigma}. The computation of $\calA(\y_0,\c_0,\f_1)$ and $\calA(\cB_0,\yB_0,\bar{\f}_1)$ yields
\begin{gather}
\calA(\y_0,\c_0,\f_1)= e^{i \arg Y_{000}+i \p (a^1_{ab}+\g_1^{ab}-\g_0^{ab})}\frac{\sqrt{\ap a^1_{ab}}}{ C_{\der Z,\s}^\t(\b_1-\a_1)} |Y_{100}| \f_1(3) \, \y_0(1) {\cdot} \c_0(2) \\
\calA(\bar{\c}_0,\bar{\y}_0,\bar{\f}_1)= e^{-i \arg Y_{000}-i \p (a^1_{ab}+\g_1^{ab}-\g_0^{ab})} \frac{\sqrt{\ap a^1_{ab}}}{ [C_{\der Z,\s}^\t]^* (1+\a_1-\b_1)} |Y_{100}| \bar{\f}_1(3) \, \cB_0(1) {\cdot} \yB_0(2)
\end{gather}
The constant in front of $|Y_{100}|$ must be one in both cases, thus the absolute value of the constant $C_{\der Z,\s}^\t$ must be
\begin{equation}
|C_{\der Z,\s}^\t(\a)|= \sqrt{\ap\,\min(\a,1-\a)}
\end{equation}
The phase $Y_{100}$ is given by the expression
\begin{equation}
\label{blablabla}
\arg Y_{100}=\arg Y_{000}+\p( \g_1^{ab}-\g_0^{ab}+a^1_{ab})-\arg C_{\der Z,\s}^\t(a^1_{ab})
\end{equation}

\subsubsection{The amplitude \texorpdfstring{$\calA(\yB_1,\y_0,\c_0,\cB_0)$}{A(psi1,psi0,chi0,chi0)}}
 
The amplitude $\calA(\yB_1,\y_0,\c_0,\cB_0)$ has same features. It reads 
\bea
\calA(\bar{\y}_1,\y_0,\c_0,\bar{\c}_0)\!&=&\!\frac{g_\textup{op}^2}{\sqrt{a^1_{bc} }} \ap\y_0(2) {\cdot}\c_0(3) \bar{\y}_1(1) {\cdot} \bar{\c}_0(4) \\
&& {\times}\! \int_0^1 dx\, x^{\ap s-1}(1-x)^{\ap t-1} \!\prod_{I=1}^3\dfrac{2\p \sqrt{\ap}}{L_{c,I} G_1^{(I)} (x)} \sum_{m_I}\frac{2\p \sqrt{\ap}m_1}{\sqrt{I_1(x)} L_{c,I}} e^{-S_\textup{Ham}^{(I)}\!(m_I,n_I)}\nn
\eea
As expected from 3-pt amplitudes \eqref{eq:chi1_BARchi0_A_vanishes} the absence of a direct coupling among $\c_1$, $\c_0$ and a gauge boson, the amplitude doesn't yield a massless pole in the $s$ channel. The leading terms of the sum over $m_1$ are given by $m_1=\pm 1$ Kaluza-Klein excitations:
\begin{equation}
\sum_{m_1} m_1 e^{-S_\textup{Ham}^{(I)}(m_I,n_I)}\xrightarrow{x\to 0} -4 i \sin \lf(2 \p \frac{f_{\c \y,1}}{L_{c,1}}\rg) x^{\frac{4\p^2 \ap}{L_{c,I}^2}}+\dots
\end{equation}
As  for the amplitudes with massless external states, we perform a Poisson resummation over the indices $m_I$ and obtain
\bea
\calA(\bar{\y}_1,\y_0,\c_0,\bar{\c}_0)&=&\frac{g_\textup{op}^2}{\sqrt{a^1_{bc}}} \ap\y_0(2) {\cdot}\c_0(3) \bar{\y}_1(1) {\cdot} \bar{\c}_0(4) \int_0^1 dx\, x^{\ap s-1}(1-x)^{\ap t-3/2-a^1_{ab}} \nn\\
&& \times \frac{G_1^{(1)}(x)}{I_1(x)} \prod_{I=1}^3\dfrac{1}{\sqrt{2\p I_I(x)}}
\sum_{\tilde{m}_I,n_I} \frac{\tilde{m}_1 L_{c,1}+f_{\c \y,1}}{2\p\sqrt{\ap}} e^{-S_\textup{Lagr}^{(I)}(\tilde{m}_I,n_I)}~~~~~~~
\eea
In the limit $x\to 1$ the leading term is the massless pole due to the chiral exchange in the $(a,b)$ sector. We have fixed all the normalizations that appear in the amplitudes and we already know the Yukawa's $Y_{010}^* Y_{000}$, thus factorization in this channel can be used to check that normalizations are consistent. The subleading terms determine the Yukawa $Y_{110}$
\begin{equation}
|Y_{110}|=|Y_{000}| \lf|\frac{2 A_{\f \y \c}^{(1)} }{\p \ap}-1\rg| \frac{1}{\sqrt{a^1_{ab} a^1_{bc}}}  \G_{1-a^1_{ab},1-a^1_{bc},-a^1_{ca}}
\end{equation}

\subsubsection{The amplitudes \texorpdfstring{$\calA(\cB_1,\c_0,\f_0,\bar{\f}_0)$}{A(chi1,chi0,phi0,phi0)} and \texorpdfstring{$\calA(\bar{\f}_1,\f_0,\c_0,\cB_0)$}{A(phi1,phi0,chi0,chi0)}}
We can obtain the Yukawas $Y_{101}$ and $Y_{011}$ from two amplitudes with massive bosons in the external states. As for the massless case, we can use Ward identities to relate these to amplitudes with only fermions. In addition to \eqref{eq:SUSY_vertex_fermions} and \eqref{eq:SUSY_vertex_bosons}, using the commutation relations
\begin{equation}
[\e {\cdot}Q,V_{\cB_1^{ji}}]=0 \qquad [\bar{\e}{\cdot} \bar{Q},V_{\cB_1^{ji}}]= V_{\d_{\bar{\e}}\bar{\f}_1^{ji}=\bar{\e}{\cdot}\cB_1^{ji}}
\end{equation}
we find 
\begin{gather}
\calA(\cB_1^{ac},\c_0^{ca},\f_0^{ab},\bar{\f}_0^{ba})=
\frac{\c_0^{ca}(2){\cdot}\c_0^{ab}(4)}{\c_0^{ab}(3){\cdot}\c_0^{ca}(2)}
\calA(\cB_1^{ac},\c_0^{ca},\c_0^{ab},\cB_0^{ba}) \\
\calA(\bar{\f}_1^{ba},\f_0^{ab},\c_0^{bc},\cB_0^{cb})=
\frac{\cB_0^{cb}(4){\cdot}\cB_0^{ab}(2)}{\cB_1^{ba}(1){\cdot}\cB_0^{cb}(4)}
\calA(\bar{\c}_1^{ba},\c_0^{ab},\c_0^{bc},\cB_0^{cb})
\end{gather}
The explicit expressions of the amplitudes read
\begin{gather}
\begin{split}
\calA(\cB_1^{ac},\c_0^{ca},\f_0^{ab},\bar{\f}_0^{ba})&= \frac{g_\textup{op}^2}{\sqrt{1+a^3_{ca}}}  \ap \bar{\c}_1^{ac}(1) \kslash_4 \c_0^{ca}(2) \int_0^1 dx\, x^{\ap s-1}(1-x)^{\ap t-5/2-a^3_{bc}} \times \\
& \times \frac{G_1^{(3)}(x)}{I_3(x)} \prod_{I=1}^3\dfrac{1}{\sqrt{2\p I_I(x)}}
\sum_{\tilde{m}_I,n_I} \frac{\tilde{m}_3 L_{a,3}+f_{\y \f,3}}{2\p\sqrt{\ap}} e^{-S_\textup{Lagr}^{(I)}(\tilde{m}_I,n_I)}
\end{split}
\\
\begin{split}
\calA(\bar{\f}_1,\f_0,\c_0,\cB_0)&=\frac{g_\textup{op}^2}{\sqrt{a^1_{ab}}} 
\ap \cB_0^{cb}(4) \kslash_2 \cB_0^{bc}(3) \int_0^1 dx\, x^{\ap s-1}(1-x)^{\ap t-5/2-a^1_{ca}} \times \\
& \times \frac{G_1^{(1)}(x)}{I_1(x)} \prod_{I=1}^3\dfrac{1}{\sqrt{2\p I_I(x)}}
\sum_{\tilde{m}_I,n_I} \frac{\tilde{m}_1 L_{b,1}+f_{\c \f,1}}{2\p\sqrt{\ap}} e^{-S_\textup{Lagr}^{(I)}(\tilde{m}_I,n_I)}
\end{split}
\end{gather}
The limit $x\to 1$ yields two new Yukawa's
\begin{align}
|Y_{011}|=& \frac{|Y_{000}|}{\sqrt{a^1_{bc}(1+a^3_{ca})}} [\G_{1-a^1_{ab},1-a^1_{bc},-a^1_{ca}} \G_{-a^3_{ab},-a^3_{bc},-a^3_{ca}}]^{1/2}
\sqrt{\frac{2  A_{\f \y \c}^{(1)}}{\p \ap} \frac{2  A_{\f \y \c}^{(3)}}{\p \ap}}\\
|Y_{101}|=& \frac{|Y_{000}|}{\sqrt{a^1_{ab}(1+a^3_{ca})}} [\G_{1-a^1_{ab},1-a^1_{bc},-a^1_{ca}} \G_{-a^3_{ab},-a^3_{bc},-a^3_{ca}}]^{1/2}
\sqrt{\frac{2  A_{\f \y \c}^{(1)}}{\p \ap} \frac{2  A_{\f \y \c}^{(3)}}{\p \ap}}
\end{align}
as well as some already computed ones that allows to constrain the phases of $Y_{100}$ and $Y_{000}$ to satisfy
 \begin{equation}
\arg Y_{000} - \arg Y_{100}=+\p( \g_1^{ab}-\g_0^{ab}) -\arg C_{\der Z,\s}^\t(a^1_{ab})
\end{equation}
Combining this with (\ref{blablabla}), one finds
\begin{equation}
\p( \g_1^{ab}-\g_0^{ab}) -\arg C_{\der Z,\s}^\t(a^1_{ab}) = - \frac{\pi}{2} a^1_{ab}
\end{equation}
As a consequence, one has 
 \begin{equation}
\arg Y_{100}= \arg Y_{000}  + \frac{\pi}{2} a^1_{ab} 
\end{equation}
and similar relations for $\arg Y_{010}$ and $\arg Y_{001}$.

\subsubsection{The amplitude \texorpdfstring{$\calA(\yB_1,\y_0,\c_0,\cB_1)$}{A(psi1,psi0,chi0,chi1)}}
The amplitude $\calA(\yB_1,\y_0,\c_0,\cB_1)$ allows us to determine the Yukawa's $Y_{111}$ and $Y_{211}$.
\bea
\calA(\bar{\y}_1,\y_0,\c_0,\bar{\c}_1)&=&\frac{g_\textup{op}^2}{\sqrt{a^1_{bc}(1+a^3_{ca})}} \ap\y_0(2){\cdot}\c_0(3) \bar{\y}_1(1) {\cdot} \bar{\c}_1(4) \int_0^1 dx\, x^{\ap s-1}(1-x)^{\ap t-1}\nn \\
&& \times  \prod_{I=1}^3\dfrac{4\p^2 \sqrt{\ap}}{L_{c,I} G_1^{(I)} (x)} \sum_{m_I,n_I}\frac{\ap m_1 m_3}{\sqrt{I_1(x)I_3(x)} L_{c,1} L_{c,3}} e^{-S_\textup{Ham}^{(I)}(m_I,n_I)}
\eea
Once again the amplitude does not expose gauge boson exchange, since the sum over the lattice forbids it. The leading contribution is given by Kaluka-Klein states
\begin{equation}
\sum_{m_1,m_3} m_1 m_3 e^{-S_\textup{H}} \xrightarrow{x\to 0} -16 \sin\lf(2 \p\frac{f_{\c \y,1}}{L_{c,1}}\rg) \sin\lf(2 \p\frac{f_{\c \y,3}}{L_{c,3}}\rg)  x^{4\p^2 \ap( 1/L_{c,1}+1/L_{c,3})}
\end{equation}
To study the $t$ channel we perform the usual Poisson resummation
\begin{equation}
\begin{split}
&\calA(\bar{\y}_1,\y_0,\c_0,\bar{\c}_1)= \frac{\ap g_\textup{op}^2}{\sqrt{a^1_{bc}(1+a^3_{ca})}} \y_0(2) {\cdot}\c_0(3) \bar{\y}_1(1) {\cdot} \bar{\c}_0(4) \int_0^1 dx\, x^{\ap s-1}(1-x)^{\ap t-5/2+a^2_{ab}} \\
& \times \frac{G_1^{(1)}(x)G_1^{(3)}(x)}{I_1(x)I_3(x)} \prod_{I=1}^3\dfrac{1}{\sqrt{2\p I_I(x)}}
\sum_{\tilde{m}_I,n_I}\frac{(\tilde{m}_1 L_{c,1}+f_{\y\c,1})(\tilde{m}_3 L_{c,3}+f_{\y\c,3})}{4\p^2 \ap} e^{-S_\textup{Lagr}^{(I)}(\tilde{m}_I,n_I)}
\end{split}
\end{equation}
Factorizing this amplitude we obtain the Yukawa's
\bea
|Y_{111}|&=&\frac{|Y_{000}|}{ \sqrt{a^1_{ab} a^1_{bc}(1+a^3_{ca})}} \G_{1-a^1_{ab},1-a^1_{bc},-a^1_{ca}} \G^{1/2}_{-a^3_{ab},-a^3_{bc},-a^3_{ca}} \lf|\frac{2 A_{\f \y \c}^{(1)}}{\p \ap}-1\rg| \sqrt{\frac{2  A_{\f \y \c}^{(3)}}{\p \ap}}\\
|Y_{211}|&=&\frac{|Y_{000}|}{\sqrt{2} a^1_{ab} \sqrt{a^1_{bc}(1+a^3_{ca})}} \G^{3/2}_{1-a^1_{ab},1-a^1_{bc},-a^1_{ca}} \G^{1/2}_{-a^3_{ab},-a^3_{bc},-a^3_{ca}} \lf|\frac{2 A_{\f \y \c}^{(1)}}{\p \ap}-3\rg| \sqrt{\frac{2 A_{\f \y \c}^{(1)}}{\p \ap} \frac{2  A_{\f \y \c}^{(3)}}{\p \ap}}\nn\\
\eea

\section{Results} \label{sect:FinResults}

In this section we summarize our results and specify our conventions, once again.

In order to have non-zero Yukawa's (R-charges) and SUSY one needs
%
%Conditions on the angles for non-zero Yukawa's (R-charges) and SUSY 
\begin{align}
a_{ab}^1+a_{bc}^1+a_{ca}^1&=0 \quad \qquad  a_{ab}^1+a_{ab}^2+a_{ab}^3=0 \\
a_{ab}^2+a_{bc}^2+a_{ca}^2&=0 \quad \qquad\,  a_{bc}^1+a_{bc}^2+a_{bc}^3=0 \\
a_{ab}^3+a_{bc}^3+a_{ca}^3&=-2 \qquad\, a_{ca}^1+a_{ca}^2+a_{ca}^3=-2
\end{align}
these lead to the combined conditions 
\begin{align}
a_{ab}^1&<2\, a_{ab}^1<a_{ab}^2<1-|a_{ab}^3| \\ 
a_{bc}^1&<2\, a_{bc}^1<a_{bc}^2<1-|a_{bc}^3| \\
|a_{ca}^1|&<2 |a_{ca}^1| < |a_{ca}^2|<|a_{ca}^3|
\end{align}
that allow for the following ranges for the angles 
\begin{align}
0&<a_{ab}^1<1/6 \qquad 0<a_{bc}^1<1/6 \quad -1/3<a_{ca}^1<0 \\
0&<a_{ab}^2<1/3 \qquad 0<a_{bc}^2<1/3 \quad -5/6<a_{ca}^2<-2/3 \\
-1/2 &<a_{ab}^3<0 \quad -1/2 <a_{bc}^3<0 \qquad \quad -1<a_{ca}^3<-5/6
\end{align}
The gauge couplings are given by 
\begin{equation}
\gYM{a}=g_\textup{op} \prod_{I=1}^3 \sqrt{\frac{\sqrt{\ap}}{L_{a,I}}}
\end{equation}
Yukawa's $Y_{ijk}$ between $\f_i$, $\y_j$ and $\c_k$, where the indices $ijk$ denote the level of the twisted field, were determined to be 
\bea
|Y_{000}|&=&\frac{ g_\textup{op}}{(2\p)^{3/4}} [\G_{1-a^1_{ab},1-a^1_{bc},-a^1_{ca}} \G_{1-a^2_{ab},1-a^2_{bc},-a^2_{ca}} \G_{-a^3_{ab},-a^3_{bc},-a^3_{ca}}]^{1/4} \times ~~~~~~~~~~~~~~\nn \\
&\quad & \times \prod_{I=1}^3 \exp\lf[- \frac{ A_{\f \y \c}^{(I)} }{ 2 \p \ap} \rg] \\
|Y_{010}|&=& \frac{|Y_{000}|}{\sqrt{a^1_{bc}}}[\G_{1-a^1_{ab},1-a^1_{bc},-a^1_{ca}}]^{1/2} \sqrt{\frac{2  A_{\f \y \c}^{(1)} }{\p\ap}}  \\
|Y_{001}|&=& \frac{|Y_{000}|}{\sqrt{1+a^3_{ca}}} [\G_{-a^3_{ab},-a^3_{bc},-a^3_{ca}}]^{1/2} \sqrt{\frac{2  A_{\f \y \c}^{(3)} }{\p\ap}}
\\
|Y_{200}|&=&  \frac{|Y_{000}|}{\sqrt{2} a^1_{ab}}\G_{1-a^1_{ab},1-a^1_{bc},-a^1_{ca}} \lf|\frac{2 A_{\f \y \c}^{(1)} }{\p\ap}-1\rg|  \\
|Y_{020}|&=& \frac{|Y_{000}|}{\sqrt{2} a^1_{bc}}\G_{1-a^1_{ab},1-a^1_{bc},-a^1_{ca}}\lf|\frac{2  A_{\f \y \c}^{(1)} }{\p\ap}-1\rg|  \\
|Y_{002}|&=& \frac{|Y_{000}|}{\sqrt{2} (1+a^3_{ca})}\G_{-a^3_{ab},-a^3_{bc},-a^3_{ca}} \lf|\frac{2  A_{\f \y \c}^{(3)} }{\p \ap}-1\rg| \\
|Y_{110}|&=&\frac{|Y_{000}|}{\sqrt{a^1_{ab} a^1_{bc}}} \G_{1-a^1_{ab},1-a^1_{bc},-a^1_{ca}} \lf|\frac{2  A_{\f \y \c}^{(1)} }{\p \ap}-1\rg| \\
|Y_{011}|&=& \frac{|Y_{000}|}{\sqrt{a^1_{bc}(1+a^3_{ca})}} [\G_{1-a^1_{ab},1-a^1_{bc},-a^1_{ca}} \G_{-a^3_{ab},-a^3_{bc},-a^3_{ca}}]^{1/2}
\sqrt{ \frac{2  A_{\f \y \c}^{(1)}}{\p \ap} \frac{2  A_{\f \y \c}^{(3)}}{\p \ap}}\\
|Y_{101}|&=& \frac{|Y_{000}|}{\sqrt{a^1_{ab}(1+a^3_{ca})}} [\G_{1-a^1_{ab},1-a^1_{bc},-a^1_{ca}} \G_{-a^3_{ab},-a^3_{bc},-a^3_{ca}}]^{1/2}
\sqrt{\frac{2  A_{\f \y \c}^{(1)}}{\p \ap} \frac{2  A_{\f \y \c}^{(3)}}{\p \ap}} \\
|Y_{111}|&=&\frac{|Y_{000}|}{ \sqrt{a^1_{ab} a^1_{bc}(1+a^3_{ca})}} \G_{1-a^1_{ab},1-a^1_{bc},-a^1_{ca}} \G^{1/2}_{-a^3_{ab},-a^3_{bc},-a^3_{ca}} \lf|\frac{2 A_{\f \y \c}^{(1)}}{\p \ap}-1\rg| \sqrt{\frac{2  A_{\f \y \c}^{(3)}}{\p \ap}}\\
|Y_{211}|&=&\frac{|Y_{000}|}{\sqrt{2} a^1_{ab} \sqrt{a^1_{bc}(1+a^3_{ca})}} \G^{3/2}_{1-a^1_{ab},1-a^1_{bc},-a^1_{ca}} \G^{1/2}_{-a^3_{ab},-a^3_{bc},-a^3_{ca}} \lf|\frac{2 A_{\f \y \c}^{(1)}}{\p \ap}-3\rg| \sqrt{\frac{2 A_{\f \y \c}^{(1)}}{\p \ap} \frac{2  A_{\f \y \c}^{(3)}}{\p \ap}}\nn\\
\eea
Twist fields normalizations are fixed to be
\begin{equation}
K_I^{c,ab}=\frac{L_{a,I}^{1/4}L_{b,I}^{5/4} L_{c,I}^{1/2}}{4\p^2 \ap}
\end{equation}
\begin{gather}
N_I^{abc}= 2^{1/6} \lf(\frac{\G_{1-\a_I,\b_I,1+\a_I-\b_I}}{2\p}\rg) ^{1/4} \sqrt{\frac{L_{a,I}}{\sqrt{\ap}}\frac{L_{b,I}}{\sqrt{\ap}}\frac{L_{c,I}}{\sqrt{\ap}}}\\
C_{\der Z,\s_\a}^{\t_\a}= \sqrt{\ap \,\min(\a,1-\a)}
\end{gather}
Finally, the overall normalization of the disk amplitude is given by
\begin{equation}
C_{D^2_a}= \frac{\prod L_{a,I}}{(2\ap)^{7/2} }
\end{equation}

\section{Conclusions and Outlook}

Notwithstanding the lack of any experimental hints for new particles beyond the Standard Model and in particular the disappearance of di-photon excess(es) at LHC, Models with low string tension, \ie $M_s \sim 5-10$ TeV, are not yet ruled out by data. Moreover, the possibility that the lightest massive string states be open string stretching between D-branes intersecting at small angles remains an interesting one. With this in mind we have reviewed the construction of BRST invariant vertex operators for such states and computed their interactions in a supersymmetric context. Due to subtleties inherent in the normalisation of vertex operators, especially those involving (un)excited twist-fields, the strategy we adopted was to extract 3-pt `Yukawa' couplings by factorisation of appropriate 4-pt amplitudes for massless states with one or two independent angles that admit an unambiguous normalisation, in that they exposed gauge boson exchange in some channel. This allowed us to re-derive the known Yukawa couplings of massless states and determine Yukawa couplings involving one `light' massive state. Later on, inserting massive external states, we determined Yukawa coupling involving two or three massive states.
We are now ready to compute partial widths and decay rates that could be eventually tested in future runs of LHC or at future accelerators.

\section*{Acknowledgements}

The authors would like to thank S. Abel, C. Angelantonj, C. Corian\'o, A. Faraggi, A. Guerrieri, E. Kiritsis, D. L\"ust, F. Morales, I. Pesando, L. Pieri, G. Pradisi for useful discussions and special thanks to R. Richter for enjoyable collaboration on these and related topics. PA would like to thank INFN Tor Vergata for hospitality during completion of this work. MB and DC would like to thank the organizers of the Corfu Summer Institute for creating a very stimulating environment. MB was partly supported by the INFN network ST$\&$FI ``String Theory and Fundamental Interactions'' and by the Uncovering Excellence Grant STaI ``String Theory and Inflation''. PA was supported by the FWF project P26731-N27.

\newpage
\appendix

\section{States and vertex operators}\label{States and vertex operators}

The vertex operators are
\bea
&& V_{A_{a_i}}^{(-1)}=C_{A_{a_i}}e^{-\f_{10}} e^{-\vf} \e_\m \y^\m e^{ikX} ~~~~~~~ \qquad V_{A_{a_i}}^{(0)}=\frac{C_{A_{a_i}}}{\sqrt{2\ap}} e^{-\f_{10}} \e_\m\lf(i \der X^\m + 2\ap k{\cdot} \y \, \y^\m \rg) e^{ikX}\qquad ~~~ \\
&& V_{\f_0=\f_0^{ab}}^{(-1)} =C_{\f_0} e^{-\f_{10}} \f_0 e^{-\vf} \s_{a^1_{a,b}} \s_{a^2_{a,b}} \s_{1+a^3_{a,b}}
e^{i[a^1_{a,b}\vf_1+a^2_{a,b}\vf_2+(a^3_{a,b}+1)\vf_3]} e^{i k X} \qquad ~~~ \\
&& V_{\y_0=\c_0^{bc}}^{(-\frac{1}{2})} =C_{\y_0} e^{-\f_{10}} \y_0^\a S_\a e^{-\frac{\vf}{2}} \s_{a^1_{b,c}} \s_{a^2_{b,c}} \s_{1+a^3_{b,c}} e^{i[(a^1_{b,c}-\frac{1}{2})\vf_1+(a^2_{b,c}-\frac{1}{2})\vf_2+(a^3_{b,c}+\frac{1}{2})\vf_3]} e^{i k X}\qquad ~~~ \\
&& V_{\bar{\y}_0 =\bar{\c}_0^{cb}}^{(-\frac{1}{2})} =C_{\y_0} e^{-\f_{10}} \bar{\y}_{0 \aD} C^\aD e^{-\frac{\vf}{2}} \s_{1-a^1_{b,c}} \s_{1-a^2_{b,c}} \s_{-a^3_{b,c}} e^{i[(-a^1_{b,c}+\frac{1}{2})\vf_1+(-a^2_{b,c}+\frac{1}{2})\vf_2+(-a^3_{b,c}-\frac{1}{2})\vf_3]} e^{i k X}\qquad ~~~ \\
&& V_{\c_0=\c_0^{ca}}^{(-\frac{1}{2})}=C_{\c_0} e^{-\f_{10}} \c_0^\a S_\a e^{-\frac{\vf}{2}} \s_{1+a^1_{c,a}} \s_{1+a^2_{c,a}} \s_{1+a^3_{c,a}} e^{i[(a^1_{c,a}+\frac{1}{2})\vf_1+(a^2_{c,a}+\frac{1}{2})\vf_2+(a^3_{c,a}+\frac{1}{2})\vf_3]} e^{i k X}\qquad ~~~ \\
&& V_{\bar{\c}_0=\bar{\c}_0^{ac}}^{(-\frac{1}{2})}=C_{\c_0} e^{-\f_{10}} \bar{\c}_{0 \aD} C^\aD e^{-\frac{\vf}{2}} \s_{-a^1_{c,a}} \s_{-a^2_{c,a}} \s_{-a^3_{c,a}} e^{i[(-a^1_{c,a}-\frac{1}{2})\vf_1+(-a^2_{c,a}-\frac{1}{2})\vf_2+(-a^3_{c,a}-\frac{1}{2})\vf_3]} e^{i k X}\qquad ~~~ \\
&& V_{\y_1=\c_1^{bc}}^{(-\frac{1}{2})}=C_{\y_1} e^{-\f_{10}} \y_1^\a S_\a e^{-\frac{\vf}{2}} \t_{a^1_{b,c}} \s_{a^2_{b,c}} \s_{1+a^3_{b,c}} e^{i[(a^1_{b,c}-\frac{1}{2})\vf_1+(a^2_{b,c}-\frac{1}{2})\vf_2+(a^3_{b,c}+\frac{1}{2})\vf_3]} e^{i k X}\qquad ~~~ \nn\\
&&~~~~~~~~~~~~+C_{\tilde\y_1} e^{-\f_{10}} \tilde\y^\dagger_{1\dot\a} C^{\dot\a} e^{-\frac{\vf}{2}} \s_{a^1_{b,c}} \s_{a^2_{b,c}} \s_{1+a^3_{b,c}} e^{i[(a^1_{b,c}+\frac{1}{2})\vf_1+(a^2_{b,c}-\frac{1}{2})\vf_2+(a^3_{b,c}+\frac{1}{2})\vf_3]} e^{i k X}\qquad ~~~ \\
&& V_{\tilde{\y}_1^\dagger =\tilde{\c}_1^{\dagger\,cb}}^{(-\frac{1}{2})}=
C_{\y_1} e^{-\f_{10}} \y^\dagger_{1\dot\a} C^{\dot\a} e^{-\frac{\vf}{2}} \t_{1-a^1_{b,c}} \s_{1-a^2_{b,c}} \s_{-a^3_{b,c}} e^{i[(-a^1_{b,c}+\frac{1}{2})\vf_1+(-a^2_{b,c}+\frac{1}{2})\vf_2+(-a^3_{b,c}-\frac{1}{2})\vf_3]} e^{i k X}\qquad ~~~ \nn\\
&&~~~~~~~~~~~~+
C_{\tilde{\y}_1} e^{-\f_{10}} \tilde{\y}_{1\a} S^\a e^{-\frac{\vf}{2}} \s_{1-a^1_{b,c}} \s_{1-a^2_{b,c}} \s_{-a^3_{b,c}} e^{i[(-a^1_{b,c}-\frac{1}{2})\vf_1+(-a^2_{b,c}+\frac{1}{2})\vf_2+(-a^3_{b,c}-\frac{1}{2})\vf_3]} e^{i k X}\qquad ~~~ \\
&& V_{\c_1=\c_1^{ca}}^{(-\frac{1}{2})}=C_{\c_1} e^{-\f_{10}} \c_1^\a S_\a e^{-\frac{\vf}{2}} \s_{1+a^1_{c,a}} \s_{1+a^2_{c,a}} \t_{1+a^3_{c,a}} e^{i[(a^1_{c,a}+\frac{1}{2})\vf_1+(a^2_{c,a}+\frac{1}{2})\vf_2+(a^3_{c,a}+\frac{1}{2})\vf_3]} e^{i k X}\qquad ~~~ \nn\\
&&~~~~~~~~~~~~+C_{\tilde\c_1} e^{-\f_{10}} \tilde\c^\dagger_{1\dot\a} C^{\dot\a} e^{-\frac{\vf}{2}} \s_{1+a^1_{c,a}} \s_{1+a^2_{c,a}} \s_{1+a^3_{c,a}} e^{i[(a^1_{c,a}+\frac{1}{2})\vf_1+(a^2_{c,a}+\frac{1}{2})\vf_2+(a^3_{c,a}-\frac{1}{2})\vf_3]} e^{i k X}\qquad ~~~ 
\eea
The vertex operators' normalizations obtained from the amplitudes are
\begin{gather}
C_{A_i}= \sqrt{2\ap}\prod_I \lf[ \frac{\ap}{L^2_{i,I}} \rg]^{1/4}  \\
%C_{A_i}= \sqrt{2\ap}\prod_I \sqrt{\frac{\sqrt{\ap}}{L_{i,I}}}  \\
C_{\c^{ij}_0}=  (\ap)^{1/4} \sqrt{2\ap} \prod_I \lf[\frac{\ap}{L_{i,I}L_{j,I}} \rg]^{1/4}  \qquad
C_{\f^{ij}_0}=  \sqrt{2\ap} \prod_I \lf[\frac{\ap}{L_{i,I}L_{j,I}} \rg]^{1/4}  \\
C_{\c_1^{ij}}=(\ap)^{1/4} \sqrt{2\ap} \prod_I \lf[\frac{\ap}{L_{i,I}L_{j,I}} \rg]^{1/4} \qquad C_{\tilde{\c}_1^{ij}}=  (\ap)^{1/4} \sqrt{2\ap} \prod_I \lf[\frac{\ap}{L_{i,I}L_{j,I}} \rg]^{1/4}
\end{gather}

\section{OPE}
In this section we summarize OPE used for the computations of correlation functions, except the twist-field correlators that will be analysed in more detail in the next appendix \ref{sect:twist_fields}.

OPE used in the computations
\begin{gather}
i \der X^\m(x_1) \,e^{i k{\cdot}X(x_2)}\sim \frac{2\ap k^\m}{x_1-x_2} e^{i k{\cdot}X(x_2)} \quad, \quad
i \der X^\m(x_1) \,i \der X^\n(x_2) \sim \frac{2\ap \h^{\m\n}}{(x_1-x_2)^2}\\
\y^\m(x_1)\, \y^\n(x_2) \sim \frac{\h^{\m\n}}{x_1-x_2}\\
\y^\m(x_1) S_\a(x_2) \sim \frac{1}{\sqrt{2}} \frac{\s^\m_{\a \dot{\a}} C^{\dot{\a}}(x_2)}{(x_1-x_2)^{1/2}} \quad, \quad 
\y_\m(x_1) C^{\dot{\a}}(x_2) \sim \frac{1}{\sqrt{2}} \frac{\bar{\s}_\m^{ \dot{\a} \a}S_\a (x_2)}{(x_1-x_2)^{1/2}} \\
\y^\m(x_1) S_\a(x_2) C_{\dot{\a}}(x_3) \sim \frac{1}{\sqrt{2}} \frac{\s^\m_{\a \dot{\a}}}{(x_1-x_2)^{1/2} (x_1-x_3)^{1/2}}
\end{gather}
Superghosts
\begin{equation}
e^{a \vf(x_1)}\, e^{b \vf(x_2)} \sim x_{12}^{-ab} e^{(a+b) \vf(x_2)} 
\end{equation}
Internal scalars $\y^I=e^{i \vf_I}$
\begin{equation}
e^{i a_I \vf_I(x_1)}\,e^{i b_I \vf_I(x_2)} \sim x_{12}^{a_I b_I} e^{(a_I+b_I) \vf(x_2)}
\end{equation}

\section{Twist field correlators} \label{sect:twist_fields}

Twist field correlators in a framework with open strings attached to D-branes can be obtained from closed strings correlators in an orbifold framework\footnote{For twist correlators involving higher excited bosonic twist fields, see \cite{David:2000yn,Lust:2004cx, Conlon:2011jq, Pesando:2011ce,Pesando:2012cx, Anastasopoulos:2013sta, Pesando:2014owa}.}. The angles $\p k/n$ associated to transformation of the group $\setZ_n$ are replaced by the angles between the branes $\p a$. Computations of bosonic twist fields correlators in an orbifold are done in \cite{Burwick:1990tu}. 

The procedure to obtain the open-string twist-field correlators requires the identification of the conformal blocks. To this end it is convenient to separate the closed-string correlator into a quantum part, which takes into account the quantum fluctuations and factories into a holomorphic and an anti-holomorphic function, and a classical part, that corresponds to a sum over all the Kaluza-Klein and winding states, weighted by the classical action for closed strings
\begin{equation}
S_\textup{cl}=\dfrac{1}{4\p \ap} \int_\setC d^2 z\,\lf(\der Z_\textup{cl} \bar{\der} \bar{Z}_\textup{cl}+ \bar{\der} Z_\textup{cl} \der \bar{Z}_\textup{cl}\rg)
\end{equation}
We distinguish closed or open-string operators putting an hat over the closed-string operators. Our starting point is the 4-pt closed-string twist-field correlator 
\begin{equation}
\begin{split}
\<{\hat\sigma}_1(z_1,\bar{z}_1)\dots {{\hat\sigma}}_4(z_4,\bar{z}_4)\>= |F(z_1,\dots,z_4)|^2 \sum_{\vec{n}_L,\vec{n}_R} c_{\vec{n}_L,\vec{n}_R} w(z)^{\ap p_L^2(n_L)/4} \bar{w}(\bar{z})^{\ap p_R^2(n_R)/4}
\end{split}
\end{equation}
where $w(z)$ is the equivalent of $q$ in the torus partition function, \ie $i\p \t(z)=\log w(z)$ the modular parameter of a `fake' torus.

The quantum part of an open-string correlator is the `square root' of the closed-string quantum correlator while the classical part is the sum over the relevant open string states
\begin{equation}
\<{\sigma}^{a,b}_1(x_1)\dots {\sigma}^{d,a}_4(x_4)\>= {\rm Tr}(T_{a,b} T_{b,c}T_{c,d} T_{d,a}) F(x_1,\dots,x_4) \sum_{\vec{k}} C^{a,c}_{\vec{k}} w(x)^{\ap p^2_{a,c}(\vec{k})}
\end{equation}
where $p^2_{a,c}(\vec{k})$ denote the generalised KK momenta carried by open strings joining the D-branes $a$ and $c$. As discussed above, the set $\{{\vec{k}} \}$ and the coefficients $C^{a,c}_{\vec{k}}$ are constrained by unitarity and planar duality.

\subsection{2-pt correlators}

Two-pt twist fields correlators in closed strings have a trivial classical part thus open (left-handed) strings are simply their holomorphic part
\begin{align}
\label{eq:twist_field_2pt}
\< \hat\s_{\a,f} (z_1)\, \hat\s_{1-\a,f}(z_2)\>&= |z_{12}|^{-2\a(1-\a)} \quad\Longrightarrow
\< \s_{\a,f} (x_1) \,\s_{1-\a,f}(x_2)\>= x_{12}^{-\a(1-\a)} \\
\< \hat\t_{\a,f} (z_1)\, \hat\s_{1-\a,f}(z_2)\>&= 0\qquad \qquad \qquad \Longrightarrow\,
\< \t_{\a,f} (x_1) \,\s_{1-\a,f}(x_2)\>= 0\\
\< \hat\t_{\a,f} (z_1)\, \hat\t_{1-\a,f}(z_2)\>&= |z_{12}|^{-2-2\a(1-\a)} \Longrightarrow\,\,
\< \t_{\a,f} (x_1)\, \t_{1-\a,f}(x_2)\>= x_{12}^{-1-\a(1-\a)}
\end{align}

\subsection{3-pt correlators}

We are particularly interested in the correlators $\<\s\,\s\,\s\>$ and $\<\t\,\s\,\s\>$. 
\begin{equation}
\label{eq:twist_field_3pt}
\begin{split}
\< \hat\s_{\a,f_1}(0)\, \hat\s_{1-\b,f_2}(1)\, \hat\s_{\b-\a,f_3}(\infty) \> &= (N^{\a,\b})^2 \sum_{\l\in \L} \exp\lf[-\frac{\sin \p \a \sin \p \b}{4 \p \ap\sin \p (\b-\a)}|f_{21}+\l|^2\rg]
\end{split}
\end{equation}
where $f_{21}=f_2-f_1$. The points $f_1$, $f_2$ and $f_3$ form a triangle where $\p \a$ is the angle at vertex $f_1$, $\p (1-\b)$ at vertex $f_2$ and $\p (\b-\a)$ at $f_3$. They are the fixed points of orbifold group transformations, for open string correlators will be interpreted as points of intersections between the branes. The sum over all the possible classical configurations is a sum over a bidimensional lattice $\L$ defined by these three points. The exponent is proportional to the area of the triangle formed by the fixed points shifted by $\l$
\begin{figure} [t]
\centering
\begin{tikzpicture}
\draw [red] (-0.2,0) -- (3.2,0);
\draw [blue] (45:{-0.2*1.4}) -- (2,2)--++(45:{0.2*1.4});
\draw [green] (3,0)++({atan2(-1,2)}:-0.2) -- (2,2)--++({atan2(-1,2)}:0.2);
\fill (0,0) circle (2 pt);
\draw (3,0) circle (2pt);
\draw (-2pt+2cm,2 cm-2pt) rectangle (2pt+2cm,2pt+2cm);
\draw (0+15 pt,0) arc (0:45:15 pt);
\node at (23:25 pt) {$\a$};
\draw (3cm-15 pt,0) arc (180:{atan2(-1,2)}:15 pt);
\node at (3 cm-30 pt,10 pt) {$1-\b$};
\node at (1.5,0-7 pt) {$c$};
\node at (1 cm-10 pt,1) {$b$};
\node at (2 cm+20 pt,1) {$a$};
\node at (-10 pt,-10 pt) {$f_1$};
\node at (3 cm+10 pt,-10 pt) {$f_2$};
\node at (2 cm,2 cm+15 pt) {$f_3$};
\end{tikzpicture}
\caption{A representation of three branes defining a triangle.}
\label{fig:triangle}
\end{figure}
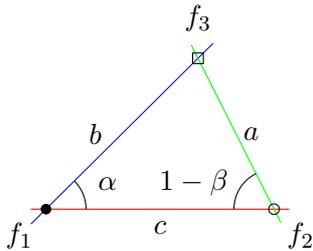
\begin{equation}
A_{f_1\,f_2\,f_3}(\l)= \frac{\sin \p \a \sin \p \b}{2 \sin \p (\b-\a)}|f_{21}+\l|^2
\end{equation}
Being a geometrical object, the area is independent of our choice of singling out the difference $f_{21}$ or for example $f_{32}$, this permutes the angles in the sines in front of the square.

Now we want to derive the open-string correlator. In this framework we have three branes $a$, $b$ and $c$. $f_1$ is an intersection between $b$ and $c$, $f_2$ is an intersection between $c$ and $a$, $f_3$ is an intersection between $a$ and $b$. The lattice $\L$ spans over two directions that are chosen between two of the three branes with lattice spacings that can be multiplies of branes' lengths. We choose the branes $b$ and $c$. The step from closed to open correlator is realized in a reduction of the sum over the (bi-dimensional) lattice $\L$ to a sum along one of the three branes considered. In the form in which we have write the correlator, the sum along the brane $b$ in any way can reproduce the area of the triangle whose vertices are the intersection points (in fact the correlator is ``$c\,$-friendly''). Hence we reduce the sum over the brane $c$. The length of the lattice spacing $\tilde{L}_c$ is fixed by requiring independence of the area from the choice of the fixed points used in the area:
\begin{equation}
\tilde{L}_c= \dfrac{I_{ab}}{\gcd(|I_{ab}|,|I_{bc}|,|I_{ca}|)} L_c
\end{equation}
where $L_c$ is the brane $c$'s length and $I_{ab}$ and similar are the number or intersections between the branes on the torus. Finally the open-string correlator is
\begin{equation}
\begin{split}
\< \s_{\a,f_1}(0)\, \s_{1-\b,f_2}(1)\, \s_{\b-\a,f_3}(\infty) \>& = \dfrac{2^{1/6}}{(2\p)^{1/4}}  \G^{1/4}_{1-\a,\b,1+\a-\b} \sqrt{\frac{L_a L_b L_c }{\ap^{3/2}} } \times \\
&\times \sum_{n_c \in \setZ} \exp\lf[- \frac{1}{2\p\ap}\frac{\sin \p \a \sin \p \b}{2\sin \p (\b-\a)} (f_{21}+n_c \tilde{L}_c)^2 \rg]
\end{split}
\end{equation}

The correlator with an excited twist field $\t$ can be computed from the correlator with one $\der Z$ and three $\s$ using the OPE between $\der Z$ and $\s$:
\begin{equation}
\< \t_{\a,f_1}(0)\, \s_{1-\b,f_2}(1)\, \s_{\b-\a,f_3}(\infty) \>=\lim_{x\to 0} \frac{x^{1-\a}}{C_{\der Z,\s}^\t} \< \der Z(x) \s_{\a,f_1}(0)\, \s_{1-\b,f_2}(1)\, \s_{\b-\a,f_3}(\infty) \>
\end{equation}
The normalization of the OPE is fixed from the amplitudes to be
\begin{equation}
\label{eq:OPE_der_Z_and_sigma}
\der Z(x_1) \, \s_\a (x_2) = C^\t_{\s,\der Z} x_{12}^{-(1-\a)} \t_\a(x_2)+\dots
\end{equation}
The correlator in the right-hand side is closely related to the unexcited correlator, in fact, as explained in details in \cite{Anastasopoulos:2013sta} the correlation function between the quantum part of $\der Z$ and the unexcited fields $\s$ vanishes.
\begin{gather}
\< \der Z(x) \s_{\a,f_1}(0)\, \s_{1-\b,f_2}(1)\, \s_{\b-\a,f_3}(\infty) \>=(\der Z)_\textup{classical}(x) \< \s_{\a,f_1}(0)\, \s_{1-\b,f_2}(1)\, \s_{\b-\a,f_3}(\infty) \>\\
(\der Z)_\textup{classical}(x)=e^{i\p(\b-\a)} x^{\a-1} (1-x)^{-\b} (f_{21}+n_c \tilde{L}_c) \frac{\G(1+\a-\b)}{\G(\a)\G(1-\b)}
\end{gather}
Hence only the classical contribution remains
\begin{equation}
\begin{split}
& \< \t_{\a,f_1}(0)\, \s_{1-\b,f_2}(1)\, \s_{\b-\a,f_3}(\infty) \>=\dfrac{2^{1/6} e^{i\p(\b-\a)}}{(2\p)^{1/4}}  \G^{1/4}_{1-\a,\b,1+\a-\b} \sqrt{\frac{L_a L_b L_c }{\ap^{3/2}} } \times \\
&\times \sum_{n_c \in \setZ} \frac{1}{C_{\der Z,\s}^\t} \frac{ \G(1+\a-\b)}{\G(\a)\G(1-\b)} (f_{21}+n_c \tilde{L}_c) \exp\lf[- \frac{1}{2\p\ap}\frac{\sin \p \a \sin \p \b}{2\sin \p (\b-\a)} (f_{21}+n_c \tilde{L}_c)^2\rg]
\end{split}
\end{equation}

\subsection{4-pt correlators}

The 4-pt correlator of unexcited twist fields has computed in \cite{Burwick:1990tu}. The result reads
\bea
\label{eq:4_pts_ssss}
&& \< \hat\s_{1-\a,f_1}(0)\, \hat\s_{\a,f_2}(z,\bar{z})\, \hat\s_{1-\b,f_3}(1)\, \hat\s_{\b,f_4}(\infty) \>=K^2  \dfrac{|z|^{-2 \a(1-\a)}(1{-}z)^{-\b(1-\a)}(1{-}\bar{z})^{-\a(1-\b)}}{I(z,\bar{z})}\nn\\
&& ~~~~~\times \sum_{\l_1,\l_2 \in \L} \exp \lf[-\dfrac{1}{4\p^2 \ap } \lf( V_{11} |u_1|^2+V_{12} \bar{u}_2 u_1+\bar{V}_{12} \bar{u}_1 u_2+V_{22}|u_2|^2 \rg)\rg]~~~~~~~~~~
\eea
where $u_1=f_{21}+\l_1$, $u_2= f_{23}+\l_2$ and $\l_i$ are vectors of the lattice $\L$. The functions $V_{ij}=V_{ij}(z,\bar{z})$ are defined as follows
\bea
V_{11}({z}{,}{\bar{z}})&=&B_1 B_2 \dfrac{\sin^2 \p \a}{4\p^2 |I|^2}\lf[|H_2|^2( G_1 \bar{H}_1+\bar{G}_1 H_1+(\a-\b) B_2 |G_2|^2)+(1\leftrightarrow 2)\rg]\\
V_{22}({z}{,}{\bar{z}})&=&\dfrac{1}{4|I|^2}\lf[G_1 G_2 (B_1 \bar{G}_2 \bar{H}_1+B_2 \bar{G}_1 \bar{H}_2)+(\text{barred}\leftrightarrow\text{unbarred})\rg]=\dfrac{\p \,I G_1 \bar{G}_2}{|I|^2}\\
V_{12}({z}{,}{\bar{z}})&=&B_1 B_2 \dfrac{\sin \p \a}{4\p |I|^2}\lf[G_2 \bar{H}_2 (2 \re [G_1 \bar{H}_1]{+}(\a{-}\b)B_2|G_1|^2){-}(\ {\leftrightarrow} 2,\text{barred}{\leftrightarrow}\text{unbarred})\rg]\nn\\
&=&B_1 B_2 \dfrac{\sin \p \a}{4\p |I|^2}\lf[2 i \im [G_1 G_2 \bar{H}_1 \bar{H}_2] +2\p(\a-\b)I G_1 \bar{G}_2\rg]
\eea
The exchange $1\leftrightarrow 2$ corresponds to the exchange $\a \leftrightarrow \b$. The functions $B_i$, $G_i$, $H_i$ and $I$ are defined as 
\bea
&& B_1=\dfrac{\G(\a)\G(1-\b)}{\G(1+\a-\b)}  \qquad \qquad B_2=\dfrac{\G(\b)\G(1-\a)}{\G(1-\a+\b)}\\
\label{eq:def_functions_G}
&& G_1=G_1(z)=\Fab{\a}{1-\b}{1}{z} \qquad  \qquad \\
&& G_2=G_2(z)=\Fab{1-\a}{\b}{1}{z} \\
&& H_1=H_1(1{-}z)=\Fab{\a}{1{-}\b}{1{+}\a{-}\b}{1{-}z} \qquad  \qquad\\
&& H_2=H_2(1{-}z)=\Fab{1{-}\a}{\b}{1{-}\a{+}\b}{1{-}z} \\
\label{eq:def_function_I}
&& I=I(z,\bar{z})=\frac{1}{2\p}\lf(B_1 \bar{H}_1 G_2+B_2 H_2 \bar{G}_1\rg)
\eea
Barred functions as $\bar{G}_1=\bar{G}_1(\bar{z})$ are the complex conjugate of the unbarred functions. The functions $B_i$, $V_{11}$ and $V_{22}$ assumes real values while $V_{12}$ is complex. There are relations connecting these functions that arise from properties of hypergeometric functions. One has
\begin{align}
G_1(z)&=\dfrac{1}{\b-\a}\lf[\dfrac{H_1(1-z)}{B_2}-(1-z)^{\b-\a}\dfrac{H_2(1-z)}{B_1}\rg]\\
G_2(z)&=\dfrac{1}{\a-\b}\lf[\dfrac{H_2(1-z)}{B_1}-(1-z)^{\a-\b}\dfrac{H_1(1-z)}{B_2}\rg]
\end{align}
also
\begin{align}
(1-z)^\a G_1(z)&=(1-z)^\b G_2(z)\\
\p(\cot\p\a-\cot \p\b)&=(\b-\a)B_1 B_2\\
G_2(z) H_1(1{-}z) B_1 - G_1(z) H_2(1{-}z) B_2 &= (\b-\a) G_1(z) G_2(z) B_1 B_2\\
I(z,\bar{z}) G_1(z) \bar{G}_2(\bar{z}) &= \bar{I} (z,\bar{z}) \bar{G}_1(\bar{z}) G_2(z)
\end{align}
The classical action can be rewritten in the form of an action for a torus worldsheet
\begin{equation}
\label{eq:classical_action_lagrangian_form}
\hat{S}(z)= \dfrac{\sin\p \a}{4\p\ap} \lf[\dfrac{1}{\t_2(z,\bar{z})} \lf|\l_2{+}f_{23}{+}\lf(\frac{\r}{2}{+} i\t_1(z,\bar{z})\rg)(f_{21}{+}\l_1)\rg|^2{+}\t_2(z,\bar{z}) |f_{21}{+}\l_1|^2\rg]
\end{equation}
where
\begin{gather}
\r= 2 \re \frac{V_{12}}{V_{22}}=\frac{\sin \p(\b-\a)}{\sin \b} \qquad \t_1(z,\bar{z})=\im \frac{V_{12}}{V_{22}} \qquad \t_2(z,\bar{z})=\dfrac{\p \sin \p \a}{V_{22}} \\
\t(z)=\t_1(z,\bar{z})+i\t_2(z,\bar{z})= i \dfrac{\sin \p \a}{2 \p} \lf[\dfrac{B_1 H_1(1{-}z)}{G_1(z)}+\dfrac{B_2 H_2(1{-}z)}{G_2(z)}\rg]
\end{gather}
$\t_1$ and $\t_2$ are real and in general depend on both $z$ and $\bar{z}$, while $\t$ is a holomorphic function. The classical action written in this way is called `lagrangian form'.

The correlator \eqref{eq:4_pts_ssss} is not in the right form to be separated in a quantum and a classical part, in fact the quantum part that appears is not in the form $F(z)\bar{F}(\bar{z})$. In order to have the right separation we perform a Poisson summation over the lattice vector $\l_2$ obtaining a new classical action
\begin{equation}
\hat{S}(z,\bar{z}){=}\dfrac{\ap \p}{\sin \p \a} \lf[\frac{i\bar{\t}(\bar{z})}{2}\lf|2\p \c{+}\frac{v_1}{4\p \ap}\rg|^2{-}\frac{i\t(z)}{2}\lf|2\p \c{-}\frac{v_1}{4\p \ap}\rg|^2\rg]-2 \p i \bar{\c} \lf[\frac{\r}{2} u_1+ f_{23} \rg]
\end{equation}
where
\begin{equation}
v_i=(1-e^{2\p i \a})u_i
\end{equation}
The action in this way is said to be in `hamiltonian form'. If we define
\begin{equation}
w(z)=\exp\lf[\frac{i \p \t(z)}{\sin\p\a}\rg]
\end{equation}
after the resummation the correlator can be written as a partition function
\begin{equation}
\label{eq:twist_field_4pt_hamiltonian}
\begin{split}
&\< \hat\s_{1-\a,f_1}(0)\, \hat\s_{\a,f_2}(z,\bar{z})\, \hat\s_{1-\b,f_3}(1)\, \hat\s_{\b,f_4}(\infty) \>= 
\lf|\frac{K^{\a,\b} z^{-\a(1-\a)}(1-z)^{-\a(1-\b)}}{G_1(z)}\rg|^2 \frac{4 \p^2 \ap}{\ell_1 \ell_2}  \\
& \sum_{\l_1\in \L,\c\in \L^*} w(z)^{\frac{\ap}{2}|2\p\c+v_1/4 \p \ap|^2}\bar{w}(\bar{z})^{\frac{\ap}{2}|2\p \c- v_1/4 \p \ap |^2} 
\exp \lf[2 \p i \bar{\c} \lf(\frac{\r}{2} u_1+ f_{23} \rg)\rg]
\end{split}
\end{equation}
where $\ell_1$ and $\ell_2$ are the lattice spacings. 

\subsubsection*{4-pt correlator for open strings (one angle un-excited fields)}

In order to find the open string correlator, the complex variable $z$ becomes the real variable $x$. Combining this reduction with the limit $\b\to\a$ we have
\begin{gather}
G_1=G_2=\Fab{\a}{1{-}\a}{1}{x}:=F_\a(z)\qquad H_1=H_2=F_\a(1{-}x)\\
B_1=B_2=\dfrac{\p}{\sin \p \a}\qquad \t_1(x)=0 \qquad \t(x)=i\dfrac{F_\a(1-z)}{F_\a(x)}:=i t(x)
\end{gather}
We note that the modular parameter has the property
\begin{equation}
\quad t(1-x)=\dfrac{1}{t(x)}
\end{equation}
The quantum part can be obtained eliminating the barred functions. The classical part is more involved. We consider two stacks of branes $b$ and $c$ intersecting at an angle $\p \a$. $f_i$ are intersecting points between $b$ and $c$. We have four integer variables, two in the sum over $\L$, that we call $n_b$ and $n_c$, and two over $\L^*$, $m_b$ and $m_c$. More explicitly
\begin{equation}
\l= n_b L_b \hat{b}+n_c L_c \hat{c} \quad \c=\frac{m_b}{L_b} \hat{b}+\frac{m_c}{L_c} \hat{c} \quad f_{21}=\frac{k_{21}}{|I_{bc}|} L_b\hat{b} \quad f_{32}=\frac{k_{32}}{|I_{bc}|} L_c\hat{c}
\end{equation}
where the variables $n$, $m$ and $k$ are integers. $\hat{b}$ and $\hat{c}$ are orthonormals. The sums must be reduced in order to have a correct open strings correlator. We expect that the classical action describe Kaluza-Klein states when the string wrap the torus along the brane direction while it describes winding states when the string wrap along a direction perpendicular to the brane. In channel s, that corresponds to the limit $x\to 0$, the states are relative to the stack $c$ thus the classical action must be:
\begin{equation}
\label{eq:channel_s_condition}
e^{-S(x)}\xrightarrow{x\to 0} f(n,m)\, x^{\frac{4\p^2\ap}{L_c^2} n^2+\frac{4\p^2 R_1^2 R_2^2}{\ap L_c^2} m^2} 
\end{equation}
We note that $L_c$ and $ R_1 R_2/L_c $ are the minimal lengths that one string must cover to wrap the torus following the parallel or perpendicular direction to the brane. In the $t$ channel (the limit $x\to 1$) the action must have the form
\begin{equation}
\label{eq:channel_t_condition}
e^{-S(x)}\xrightarrow{x\to 1} g(\tilde{n},\tilde{m}) (1-x)^{\frac{4\p^2 \ap}{L_b^2} \tilde{n}^2+\frac{4\p^2 R_1^2 R_2^2}{\ap L_b^2} \tilde{m}^2}
\end{equation}
(the tilde is due to the Poisson resummation that must be done). The classical action for closed strings in the limit $\b=\a$ is simpler
\begin{equation}
\hat{S}(z=x,\bar{z}=x)=\frac{\p \t(x)}{\sin \p \a} \lf(\ap |2\p\c|^2 +\frac{\sin^2 \p \a}{4 \p^2 \ap} |\l+f_{21}|^2\rg)+2\p i \bar{\c} f_{32}
\end{equation}
In limit $x\to 0$ the classical action assumes the form\footnote{The angle $\a$ are linked to the number of intersections by the relation $\sin \p \a = I_{bc} \frac{4\p^2 R_1 R_2}{L_b L_c}$ with $R_i$ the radii of the torus.}
\begin{equation}
e^{-S(x)} \xrightarrow{x\to 0} e^{-2\p i m_c \frac{k_{32}}{|I_{bc}|}}\, x^{\ap \frac{4 \p^2 m_c^2}{L_c^2}+\frac{4 \p^2 R_1^2 R_2^2}{ \ap L_c^2} (n_b |I_{bc}|+k_{21})^2+\frac{\sin^2 \p \a}{4 \p^2 \ap L_c^2}n_c^2+\ap \frac{4 \p^2 m_b^2}{L_b^2}} 
\end{equation}
Comparing this formula with \eqref{eq:channel_s_condition}, in order to have terms independent from the brane $b$ the only way is to fix $n_c=0$ and $m_b=0$. The action reduces to
\begin{equation}
\label{eq:action_open_two_angles}
S(x)=\frac{\p t(x)}{\sin \p \a} \lf(\frac{4\p^2 \ap}{L_c^2} m_c^2+\frac{4\p^2 R_1^2 R_2^2}{\ap L_c^2} (n_b |I_{bc}|+k_{21})^2\rg)+2\p i m_c \frac{k_{32}}{|I_{bc}|}
\end{equation}
Using Poisson summations on the remain variables, $n_b$ and $m_c$, and taking $x\to 1$ we find a match with \eqref{eq:channel_t_condition}. Finally we rewrite the complete correlator as
\begin{equation}
\begin{split}
&\< \s_{1-\a,f_1}(0)\, \s_{\a,f_2}(x)\, \s_{1-\a,f_3}(1)\, \s_{\a,f_4}(\infty) \>= \dfrac{K^{c,b} [x(1-x)]^{-\a(1-\a)}}{ F_\a(x)}\times\\
&\times \sum_{n_b,m_c} \frac{4 \p^2 \ap }{ L_b L_c} \exp \lf[-\frac{\p t(x)}{\sin \p \a} \lf(\frac{4\p^2 \ap}{ L_c^2} m_c^2+\frac{\sin^2 \p \a}{4\p^2\ap} (n_b L_b +f_{21})^2\rg)-2\p i m_c \frac{f_{32}}{L_c}\rg]
\end{split}
\end{equation}

\subsubsection*{4-pt correlator for open strings (two angles and un-excited fields)}

The case with two independent angles is very similar to one angle case. The classical action has one more term
\begin{equation}
\hat{S}(z)=\frac{\p \t(z)}{\sin \p \a} \lf(\ap |2\p\c|^2+\frac{\sin^2 \p \a}{4\p^2 \ap} |\l+f_{21}|^2\rg)-2 \p i \bar{\c} \lf(\frac{\r}{2} u_1+ f_{23} \rg)
\end{equation}
The vectors $f_i$ are proportional to the relative length with a real coefficient instead of an integer as in one angle case. The steps to obtain the open strings correlator are the same thus we report only the result
\begin{equation}
\begin{split}
&\< \s_{1-\a,f_1}(0)\, \s_{\a,f_2}(x)\, \s_{1-\b,f_3}(1)\, \s_{\b,f_4}(\infty) \>=
\dfrac{K^{c,ab} x^{-\a(1-\a)}(1-x)^{-\a(1-\b)}}{G_1 (x)}\times\\
&\times \sum_{n,m} \frac{4\p^2 \ap}{L_b L_c} \exp \lf[{-}\frac{\p t(x)}{\sin \p \a} \lf(\frac{4\p^2 \ap}{L_c^2} m^2+\frac{\sin^2 \p \a}{4\p^2 \ap} \lf(\dfrac{|I_{ca}|}{\gcd(|I_{bc}|,|I_{ca}|)} n L_b{+}f_{21}\rg)^2\rg)\!{-}2\p i m \frac{f_{32}}{L_c} \rg]
\end{split}
\end{equation}

\subsubsection*{4-pt correlator for open strings (two angles and one excited field)}

The correlator with an excited twist field $\t$ can be computed as was done for the 3-pt correlator:
\begin{equation}
\begin{split}
\< \t_{1-\a,f_1}(0)\, \s_{\a,f_2}(x)\, &\s_{1-\b,f_3}(1)\, \s_{\b,f_4}(\infty) \>=\\
&=\lim_{y\to 0} \frac{y^{1-\a}}{C_{\der Z,\s}^\t} \< \der Z(y) \s_{1-\a,f_1}(0)\, \s_{\a,f_2}(x)\, \s_{1-\b,f_3}(1)\, \s_{\b,f_4}(\infty) \>
\end{split}
\end{equation}
However the correlator in the right hand is similar to the correlator without excited twist fields, $(\der Z)_\textup{qu}$ does not correlate with four $\s$ thus remains the classical contribution
\begin{equation}
(\der Z)_\textup{classical}(y)= y^{-(1-\a)} \lf[e^{i\p \a} \frac{\sin \p \a}{\p} B_1 H_1(1-x) u_1 + G_1(x) u_2 \rg]
\end{equation}
This expression is derived from an action in the `lagrangian' form, thus we combine it with a correlator in the same form and obtain:
\begin{equation}
\begin{split}
&\< \s_{1-\a,f_1}(0)\, \s_{\a,f_2}(x)\, \s_{1-\b,f_3}(1)\, \s_{\b,f_4}(\infty) \>=
\dfrac{K^{c,ab} x^{-\a(1-\a)}(1-x)^{\a \b-(\a+\b)/2}}{\sqrt{I(x)}}\times\\
& \times \sum_{n,\tilde{m}} \frac{2\p \sqrt{\ap}}{L_b} \lf[e^{i\p \a}\frac{\sin \p \a}{\p} B_1 H_1(1-x) (L_b n +f_{21})+G_1(x)(L_c \tilde{m}+f_{32}) \rg] \times \\
& \times \exp \lf[-\frac{\sin \p \a}{4 \p \ap} \lf(t(x) \lf(\dfrac{|I_{ca}|}{\gcd(|I_{bc}|,|I_{ca}|)} n L_b+f_{21}\rg)^2+\frac{1}{t(x)} \lf(L_c \tilde{m}+f_{32}\rg)^2\rg)\rg]
\end{split}
\end{equation}

\section{Useful relations}
We here collect some definitions
\begin{gather}
\r=\dfrac{\sin\p(\b-\a)}{\sin \p\b}\\
\G_{a,b,c}=\dfrac{\G(a)\G(b)\G(c)}{\G(1-a)\G(1-b)\G(1-c)}\\
t(x)=\t_2(x)=\sin \p \a \frac{I(x)}{G_1(x) G_2(x)}
\end{gather}
as well as Poisson resummation formulae
\begin{gather}
\sum_{n} e^{-\p a (n+b)^2}= \dfrac{1}{\sqrt{a}}\sum_m e^{-\p \frac{m^2}{a}+2\p i m b} \\
\sum_n n e^{-\p a n^2+\p b n}=\sum_m \frac{-i}{a^{3/2}} \lf(m+\frac{ib}{2}\rg) e^{-\frac{\p}{a}(m+ib/2)^2}\\
\sum_{\l\in \L} e^{-\p(\l+x)^+ A (\l+x)}= \dfrac{1}{\sqrt{|\det A|}} \sum_{\c\in \L^*} e^{-\p \c^+ A^{-1} \c+2\p i \c^+ x}\\
\sum_{\l\in \L} \l e^{-\p(\l+x)^+ A (\l+x)}= \dfrac{1}{\sqrt{|\det A|}} \sum_{\c\in \L^*} (iA^{-1} \c - x)e^{-\p \c^+ A^{-1} \c+2\p i \c^+ x}
\end{gather}
In order to factorize 4-pt amplitudes onto 3-pt ones, one needs the limiting behaviour of hypergeometric functions as well as of other structures.
Limits with one independent angle produce
\begin{gather}
F_\a(x)\xrightarrow{x\to 0}1+\dots \quad F_\a(1-x)\xrightarrow{x\to 0} -\log x +2\y(1)-\y(\a)-\y(1-\a):=\log(\d_\a/x)\\
t(x)\xrightarrow{x\to 0} -\frac{\sin \p \a}{\p} \log\frac{x}{\d_\a}\qquad \frac{1}{t(x)}=t(1-x)\xrightarrow{x\to 1} -\frac{\sin \p \a}{\p} \log\frac{1-x}{\d_\a}
\end{gather}
Limits with two independent angles ($x\to 0$ and $\b>\a$) produce
\begin{gather}
G_i(x)\xrightarrow{x\to 0} 1\qquad H_1(1-x) \xrightarrow{x\to 0} \frac{1}{B_1}\lf(-\log x +2\y(1)-\y(\a)-\y(1-\b)\rg)\\
t(x) \xrightarrow{x\to 0} \dfrac{\sin\p \a}{2\p}\lf[{-}2\log x{+}4 \y(1){-}\y(\a){-}\y(1{-}\a){-}\y(\b){-}\y(1{-}\b) \rg]{=}{-}\dfrac{\sin\p \a}{\p} \log \frac{x}{\sqrt{\d_\a \d_\b}}
\end{gather}
Limits with two independent angles ($x\to 1$ and $\b>\a$) produce
\bea
&& G_1(x)\xrightarrow{x\to 1}\dfrac{\G(\b-\a)}{\G(\b)\G(1-\a)}\lf[1-(1-x)^{\b-\a} \G_{1-\a,\b,1+\a-\b}\rg]+\dots \qquad \\
&& H_i(1-x)\xrightarrow{x\to 1} 1+\dots\\
&& t(x)\xrightarrow{x\to 1} \frac{\r}{2} \lf[1+\frac{2 \G_{1-\a,\b,1+\a-\b}}{\b-\a} (1-x)^{\b-\a}+\frac{2\G^2_{1-\a,\b,1+\a-\b} }{(\b-\a)^2} (1-x)^{2(\b-\a)}\rg]+\dots\nn\\
\\
&& I(x)\xrightarrow{x\to 1} \frac{(1-x)^{\a-\b}}{2\p \G_{1-\a,\b,1+\a-\b}} \lf[1-\frac{\G_{1-\a,\b,1+\a-\b}^2}{(\b-\a)^2}(1-x)^{2(\b-\a)} \rg]+\dots\\
&& e^{-S(\tilde{m}_I=0,n_I=0)}\xrightarrow{x\to1} \exp\lf[- \frac{ A_{\f \y \c}^{(I)}}{2\p \ap} \rg] \lf[
1+\frac{\G_{1-\a,\b,1+\a-\b}}{\b-\a}\frac{2 A_{\f \y \c}^{(I)} }{\p \ap}  (1-x)^{\b-\a} +\rg. \\
&& ~~~~~~~~~~~~~~~~~~~~~~~~~~~~~ +\lf. \frac{\G_{1-\a,\b,1+\a-\b}^2}{(\b-\a)^2}\frac{2 A_{\f \y \c}^{(I)} }{\p \ap}  \lf(\frac{ A_{\f \y \c}^{(I)} }{\p\ap}-1\rg)(1-x)^{2(\b-\a)}\rg]+\dots \nn
\eea

\clearpage %\nocite{*}

\bibliographystyle{JHEP}
%\bibliography{references_TwistDecay}

\providecommand{\href}[2]{#2}\begingroup\raggedright\endgroup

\end{document}